\documentclass{aa}

\usepackage{graphicx,txfonts}
\usepackage{natbib}
\bibpunct{(}{)}{;}{a}{}{,} 

\def\nng{n_{{\rm g}n}}
\def\npg{n_{{\rm g}p}}

\begin{document}

\title{
The proto-neutron star inner crust in the liquid phase}

\author{H. Dinh Thi\inst{1}, 
A.~F. Fantina\inst{2},
F. Gulminelli\inst{1}}

\institute{Normandie Univ., ENSICAEN, UNICAEN, CNRS/IN2P3, LPC Caen, 14000 Caen, France \\
\email{dinh@lpccaen.in2p3.fr}
\and Grand Acc\'el\'erateur National d'Ions Lourds (GANIL), CEA/DRF - CNRS/IN2P3, Boulevard Henri Becquerel, 14076 Caen, France
}

\date{Received xxx Accepted xxx}

\abstract  
{ 
The crust of a neutron star is known to melt at a temperature that increases with increasing matter density, up to about $ 10^{10}$~K. At such high temperatures and beyond, the crustal ions are put into collective motion and the associated entropy contribution can affect both the thermodynamic properties and the composition of matter.
}
{ 
We studied the importance of this effect in different thermodynamic conditions relevant to the  inner crust of the proto-neutron star, both at beta equilibrium and in the fixed-proton-fraction regime.
}
{ 
To this aim, we solved the hydrodynamic equations for an ion moving in an incompressible, irrotational, and non-viscous fluid, with different boundary conditions, thus leading to different prescriptions for the ion effective mass.
We then employed a compressible liquid-drop approach in the one-component plasma approximation, including the renormalisation of the ion mass to account for the influence of the surrounding medium.
}
{ 
We show that the cluster size is determined by the competition between the ion centre-of-mass motion and the interface properties, namely the Coulomb, surface, and curvature energies.
In particular, including the translational free energy in the minimisation procedure can significantly reduce the optimal number of nucleons in the clusters and lead to an early dissolution of clusters in dense beta-equilibrated matter.
On the other hand, we find that the impact of translational motion is reduced in scenarios where the proton fraction is assumed constant and is almost negligible on the inner-crust equation of state.
}
{ 
Our results show that the translational degrees of freedom affect the equilibrium composition of beta-equilibrated matter and the density and pressure of the crust-core transition in a non-negligible way, highlighting the importance of its inclusion when modelling the finite-temperature inner crust of the (proto-)neutron star. 
}


\titlerunning{The proto-neutron star inner crust in the liquid phase}
\authorrunning{Dinh Thi et al.}

\maketitle

\section{Introduction}
\label{sec:intro}

The inner crust of a neutron star (NS), a region of $\approx 1$~km thickness laying outside the homogeneous nuclear-matter core of the star, is known to play a key role in several phenomena related to the physics of the whole star, particularly the cooling process \citep{Newton2013, Horowitz2015,Lin2020} and its transport properties (see e.g. \citet{Schmitt2018} for a review).
The inner crust is made of a periodic lattice of extremely exotic ions embedded in a highly relativistic electron gas; the nuclei are so neutron rich that the neutron eigenstates lay high in the continuum and constitute a superfluid neutron gas in which the nuclei are immersed.
Because of this peculiar structure, the most realistic modelling of the NS inner crust available today is based on microscopic self-consistent Hartree-Fock (HF), Hartree-Fock+BCS, or Hartree-Fock-Bogoliubov (HFB) calculations (see e.g. \citet{MagHen2002, Baldo2007, GogMut2007, Grill2011, Shelley2020}),
or on the extended Thomas-Fermi plus Strutinski Integral method based on accurately calibrated functionals (see e.g. \citet{Pearson2018, Pearson2020, Shelley2021, pearson2022}).
These approaches allow a self-consistent treatment of the nucleonic states and the inclusion of pairing and shell effects.
Some of the aforementioned works also allowed for extension to include the description of strongly deformed clusters at the edge of the crust-core transition, known as `pasta' phases.
These microscopic calculations usually employ effective functionals optimised on both experimental data, such as nuclear masses, and ab initio calculations of nuclear matter, and the residual model dependence can be accurately quantified with statistical tools \citep{Pastore2017}. 

In finite-temperature phenomena, such as proto-NS cooling and NS mergers, where the temperature can easily exceed values of several $10^{10}$~K, the crust is expected to be made of a liquid multi-component plasma composed of different nuclear species in strong interaction with the neutron (and, at sufficiently high temperature, even the proton) gas (see e.g. \citet{Oertel2017, Burgio2018} for a review). 
In this regime, the Wigner-Seitz (WS) cell is not related to any lattice structure but only represents the smallest charge-neutral volume scale, where the ion translational motion should be accounted for. 
In particular, in the low-density, high-temperature nuclear statistical equilibrium limit, it is the competition between the translational and the mass term that determines the ion abundances, which are given by the stationary solution of the Saha equations (see e.g. \citet{Hillebrandt1984}).
This multi-component structure of the NS matter necessitates beyond-mean-field methods at a mesoscopic scale, which become computationally demanding for a full density functional treatment.

If we limit ourselves to moderate temperatures close to the crystallisation transition, $k_{\rm B} T_{\rm m}\approx 0.1-1$~MeV \citep{hpy2007}, with $k_{\rm B}$ being the Boltzmann constant, \citet{Carreau_MCP} showed that the cluster distribution is relatively narrow at least at the lowest densities in the inner crust, meaning that the one-component plasma (OCP) approximation can be justified \citep{Burrows1984}.
Still, in this temperature regime, the relative collective motion of the cluster and the nucleon fluid has to be included.
Indeed, the melting process is mainly due to the competition between vibrational and translational degrees of freedom \citep{PC2000, hpy2007, MedCum2010, Fantina2020, Carreau2020}, and these latter become more and more important with increasing temperature. 
However, accounting for the collective motion of the clusters in variational OCP approaches based on nucleonic degrees of freedom, such as the microscopic HFB, is not a straightforward task, particularly from a computational point of view.

Because of these difficulties and the intrinsic complexity of multi-component plasma calculations, since the pioneering work by \citet{lattimer1985}, most approaches to the finite-temperature crust employed the OCP approximation with cluster ---rather than nucleon--- degrees of freedom (see e.g. \citet{Avancini2009, Avancini2017, Fantina2020, Carreau2020}). 
This same approximation is employed in most so-called `general-purpose' equations of state currently used in supernova and merger simulations (\citet{lattimer1991, Shen2011, GShen, Schneider}; see also \citet{Oertel2017, Burgio2018} for a review).
In these formalisms, it is, in principle, straightforward to include translational degrees of freedom for the clusters. 
However, the translational term is often neglected \citep{Avancini2009, Shen2011, Avancini2017} or excluded from the variational equation with respect to the cluster size \citep{lattimer1991, Schneider}.
  
Indeed, evaluating the mobility of an ion in a fixed nucleon background is not a simple problem. 
Many works have been devoted to a similar issue, namely the mobility of superfluid neutrons in the ion rest frame in the case of a collective rotation of the solid crust. 
In this case, a non-dissipative drag due to Bragg scattering was predicted, the so-called `entrainment' effect \citep{Chamel2005, Chamel2006,Chamel2012,Chamel2017}. 
This effect is due to the complex structure of the Fermi surface, originating from the neutron band structure generated by the periodic ion potential. 
The momentum transfer to the lattice of dripped neutrons lying in the energy bands imposed by the Bloch boundary conditions is found to lead to effective binding of the neutrons to the periodic structure. 
This would correspond to an increase in the effective mass of the clusters, even if the direction of the momentum transfer is still a matter of debate, with recent works reporting an `anti-entrainement' with an increased neutron mobility and consequently a reduction of the ion effective mass \citep{Kashiwaba2019,Sekizawa2022}. 

However, in the liquid phase, the situation is very different. 
The motion of the different clusters is fully uncorrelated, collective lattice effects are absent, and the Fermi surface of dripped neutrons is essentially spherical. 
The liquid phase problem closely resembles that of the fluid-dynamic flow of neutrons around impurities, as treated by different authors \citep{Epstein1988, Magierski2004, Magierski2004b, Martin2016}. 
In the hypothesis of an irrotational and incompressible flow, the local hydrodynamic equations admit an analytical solution, leading to a systematic reduction of the cluster effective mass. 
This can be explained given that the cluster, considered as a finite portion of nuclear matter at higher density with respect to the nucleon `gas', is assumed to be permeable to the surrounding nuclear fluid. 
Therefore, in the frame of the cluster, the unbound neutrons counter-flow and the cluster effectively moves with a reduced speed.
According to the hydrodynamical description, the ion inertia can be increased only if the nucleus is considered as an impermeable solid obstacle~\citep{Sedrakian1996} displacing the neutron fluid in its motion and therefore leading to an effective drag, a picture that is difficult to reconcile with the microscopic treatment of nuclear dynamics~\citep{Negele1982, Jin2021}.

In this paper, we carry out a study of the effect of the inclusion of translational degrees of freedom on the equation of state and the composition of matter in the liquid phase in the density regime relevant to the inner crust of the (proto-)NS. 
We employed the formalism of \citet{Carreau2019}, \citet{Carreau2020}, \citet{dinh2021}, \citet{dinhEPJA21}, and \citet{Grams2022} that, although not as microscopic as a full density functional treatment, was recently shown to provide results in good agreement with extended Thomas-Fermi calculations both at zero \citep{Grams_arxiv} and finite temperature \citep{Carreau2020}.
This approach is extended here to include the possible presence of proton drip and to compute the exact numerical calculation of the Fermi integrals beyond the Sommerfeld approximation previously used in \citet{Carreau2020}. 
Our calculations were performed in two different conditions: (i) the beta-equilibrium and (ii) the fixed-proton-fraction case.
The former situation is achieved in late (proto-)NS cooling stages, after tens of seconds from the birth of the hot proto-NS \citep{Burrows1986, Prakash1997, Yakovlev2004, Page2006}.
The second condition is of interest for astrophysical scenarios, either where the total proton fraction can be assumed to be roughly constant, as in some regions of the supernova collapsing core in the neutrino-trapped regime (e.g. \citet{Sato1975, Bruenn1985}; see also \citet{Lieben2005}), or for the computation of general-purpose equations of state, where calculations are performed for given (fixed) values of density $n_B$, temperature $T$, and total proton (or electron) fraction $Y_p^{\rm tot}$ (see e.g. \citet{lattimer1991, GShen, Shen2011, Schneider}; see also \citet{Oertel2017, Compose} for an overview of general-purpose equation-of-state computation).
For this reasons, several studies have been devoted to calculations at finite temperature and fixed proton fraction ($Y_p^{\rm tot} \approx 0.1 - 0.5$), particularly for conditions relevant to supernova matter (e.g. \citet{Avancini2009, Newton2009, Pais2012, Pais2014, Avancini2017, Ji2020}).

The paper is organised as follows: In Sect.~\ref{sec:hydro} we review the standard hydrodynamic calculation of ion motion and derive the expressions for the  effective mass of an ion, treating the ion as a hard impenetrable sphere in Sect.~\ref{sec:hydro-hard-sphere} or as a permeable sphere in Sect.~\ref{sec:hydro-sphere}.
We subsequently introduce our formalism for the proto-NS inner crust in Sect.~\ref{sec:formalism}.
After outlining the expression of the bulk, Coulomb, surface, and curvature free-energy terms in Sects.~\ref{sec:nucfunc} and \ref{sec:finsize}, special attention is given to the derivation of the translational free energy in Sect.~\ref{sec:trans}.
In particular, we discuss the different modifications to the ideal-gas expression that have to be introduced in the dense medium.
Results at beta equilibrium are presented in Sect.~\ref{sec:result_betaequi} while those in the fixed-proton-fraction scenario are shown in Sect.~\ref{sec:result_fix_yp}.
Conclusions are given in Sect.~\ref{sec:conclusion}.

\section{Hydrodynamic approach of the ion motion}
\label{sec:hydro}

In the liquid phase we are interested in, the translational degrees of freedom of the ions should be accounted for. 
As discussed in Sect.~\ref{sec:intro}, calculating the ion mobility in a nucleon background is not straightforward.
If the effect of the nucleon background is neglected, the free energy corresponding to the translational motion reduces to the standard expression of an ideal gas \citep{hpy2007} (see Sect.~\ref{sec:trans}).
However, densities in the NS crust are relatively high, particularly at the bottom of the crust, and therefore the in-medium effects cannot be ignored.
In particular, the cluster bare mass $M_i = (A-Z) m_n + Z m_p$ ---where $A$ and $Z$ are the ion mass and proton number, respectively, and $m_n$ ($m_p$) is the neutron (proton) mass--- should be replaced by an effective mass $M_i^\star$ accounting for the modification of the free flow in a nuclear medium.

In this section, we review the derivation of the effective mass of the ion moving in an incompressible and irrotational fluid, within an ideal hydrodynamic approach, as already proposed in the literature for a different application~\citep{Epstein1988, Sedrakian1996, Magierski2004, Magierski2004b, Martin2016, Chamel2017}.
We present the solutions to the hydrodynamic equations with two different boundary conditions, which correspond to the assumptions that the ion is (i) an impenetrable hard sphere, and (ii) a permeable sphere.

\begin{figure}
    \centering
    \includegraphics[scale = 0.45]{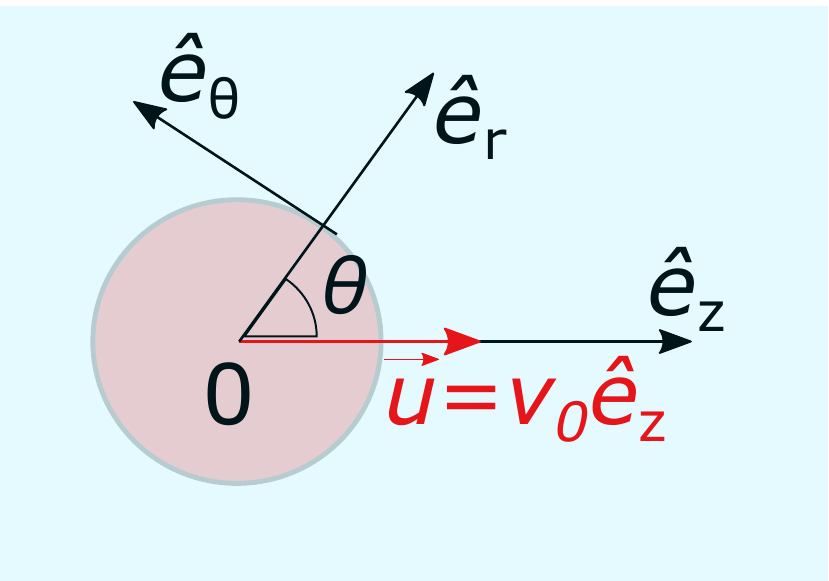}
    \caption{Sketch of the motion of the ion (red sphere) in a uniform background, with the considered system of reference.}
\label{fig:ion-motion}
\end{figure}

Let us start by considering an ion ---with density $n(r) = n_i$ for $r<r_N$, where $r_N$ is the ion radius (therefore $V_N = 4 \pi r_N^3/3$ is the ion volume)--- moving in a uniform background with density $n(r)=n_{\rm g} = \nng + \npg$ for $r>r_N$, where $\nng$ ($\npg$) is the neutron (proton) component.
The ion motion is sketched in Fig.~\ref{fig:ion-motion}: the ion centre-of-mass velocity $\vec{u}$ is directed along the $z$ axis, with the ion centre of mass coinciding with the origin of the system of reference, that is, $\vec{u} = v_0 \vec{\hat{e}}_z$. 
As nuclear matter both inside and outside the ion is assumed to be incompressible and irrotational, there exists a velocity potential solution of the Laplace equation $\nabla^2\Phi(\vec{r})=0$  with $\vec{v}=\nabla \Phi$, such that
$\Phi(\vec{r}) = \Phi_{\rm in} (\vec{r}) + \Phi_{\rm out} (\vec{r})$,
with $\Phi_{\rm in}(\vec{r})$ and $\Phi_{\rm out}(\vec{r})$ being the potential inside ($r<r_N$) and outside ($r>r_N$) the cluster, respectively. 
 
Because of the azimuthal symmetry of the system, $\Phi(\vec{r}) = \Phi(r,\theta)$ (see Fig.~\ref{fig:ion-motion}), the standard separable ansatz, $\Phi(r,\theta) = R(r) \Theta(\theta)$, leads to the following general solution of the Laplace equation, 
\begin{equation}
    \Phi_{\rm k}(r,\theta) = \sum_{l=0}^\infty P_l(\cos \theta) \left( \mathcal{A}^{(l)}_{\rm k} r^l + \frac{\mathcal{B}^{(l)}_{\rm k}}{r^{l+1}} \right) \ ,
\label{eq:poisson-sol}
\end{equation}
where ${\rm k}={\rm \{in, out\}}$, $P_l$ is the Legendre polynomial, and the coefficients of the series have to be determined from the boundary conditions.
 In the ion reference frame moving at speed $\vec{u}$ with respect to the crust, the general condition that the nucleon fluid should be asymptotically unaffected by the ion motion, that is 
$\lim_{r\to\infty}\vec{v}(\vec{r})= -  \vec{u} $, implies $l=1$ in Eq.~(\ref{eq:poisson-sol}) for the outer region, leading to:
\begin{equation}
    \Phi_{\rm out}(r,\theta) = \left( \mathcal{A}_{\rm out} r + \frac{\mathcal{B}_{\rm out}}{r^{2}} \right) \cos \theta \ ,
    \label{eq:poisson-sol-l1}
\end{equation}
independently of the conditions at the surface between the ion and the fluid.

\subsection{Ion as an impenetrable hard sphere}
\label{sec:hydro-hard-sphere}

Let us first consider the ion as an impenetrable hard sphere.
In this case, the velocity potential $\Phi_{\rm in}$ is fully determined by the condition that the velocity field is constant inside the ion, $\vec{v}(r<r_N,\theta)=\vec{u}$.
Concerning $\Phi_{\rm out}$, the constants $\mathcal{A}_{\rm out}$ and $\mathcal{B}_{\rm out}$ in Eq.~(\ref{eq:poisson-sol-l1}) are determined by imposing that in the crust reference frame, the nucleon fluid is at rest far from the ion, $\vec{v}(r\to\infty,\theta)=\vec{0}$, and the sphere is impenetrable, meaning that the ion `pushes' the external nucleon fluid along in the same direction of motion, that is, $v_r(r\to r_N,\theta)=u_r$, where $v_r$ and $u_r$ are the radial components of $\vec{v}$ and $\vec{u}$, respectively.
This leads to the following expression for the potential field $\Phi$:
\begin{equation}
\Phi(r,\theta) = 
\begin{cases}
    v_0 r \cos \theta \  & \text{if } r<r_N \ ,\\
    -\frac{v_0 r_N^3}{2 r^2} \cos \theta  & \text{if } r>r_N \ .
\end{cases}
\label{eq:poisson-sol-hard-sphere}
\end{equation}
The kinetic energy associated to the flow can therefore be calculated as 
\begin{eqnarray}
E_k(\vec{v})&=&\frac{1}{2} \int_{V} d^3 \vec{r} \rho_b(r) |\nabla \Phi|^2 \label{eq:ek-int} \nonumber \\
&=& \frac{1}{2} \int_{V} d^3 \vec{r} \rho_b(r) \nabla \cdot \left (\Phi \nabla \Phi \right ) \ , 
\end{eqnarray}
where $\rho_b(r)$ is the local (baryonic) mass density, and the integral is extended to the total volume.
Applying the divergence theorem, the kinetic energy reads
\begin{eqnarray}
    E_k &=& \frac{2}{3} \pi v_0^2 r_N^3 \left( \rho_{b,i} + \frac{1}{2} \rho_{b,{\rm g}} \right) \nonumber \\
     &=& \frac{1}{2} M_i \left( 1 + \frac{1}{2} \frac{\rho_{b,{\rm g}}}{\rho_{b,i}} \right) v_0^2 = \frac{1}{2} M^\star_i v_0^2 \ ,
     \label{eq:ek-hard-sphere}
\end{eqnarray}
with $\rho_{b,i} = M_i/V_N$ being the mass density of the ion and $\rho_{b,{\rm g}} = m_p \npg + m_n \nng$ being that of the outside fluid.
Defining $\gamma \equiv \rho_{b,{\rm g}} / \rho_{b,i}$, we find 
\begin{equation}
    M_i^\star = M_i \left( 1 + \frac{1}{2} \gamma \right) \ ,
\label{eq:m*-hard-sphere}
    \end{equation}
which reduces to $M_i^\star = M_i$ for nuclei in vacuum ($\gamma=0$).
Therefore, modelling the ion as an impenetrable hard sphere leads to an increase in the ion effective mass.
Making the further approximation that the neutron and proton mass be equal, one retrieves Eq.~(2.43) of \citet{lattimer1985} (although in the latter work the outside medium also includes alpha particles).
Additionally, if there are no dripped protons, and therefore $n_{\rm g} = \nng$ and $\gamma = \nng / n_i$, one recovers the picture of the neutron fluid flowing around the cluster, as mentioned in \citet{Sedrakian1996}.

\subsection{Ion as a permeable sphere}
\label{sec:hydro-sphere}

From the microscopic point of view, the cluster and the dripped nucleons are portions of the same fluid (nuclear matter) with different densities and proton-to-neutron ratios. One therefore expects that in the ion reference frame (at least a portion of) the nucleons in the cluster participate in the flow of the (external) fluid.
This latter situation was studied by \citet{Magierski2004, Magierski2004b}, who modelled the ion and the dripped neutrons as a single fluid and found that, as opposed to the hard-sphere image, a uniform flow of neutrons penetrates through the cluster.
This latter is also an extreme picture, because we  only expect the nucleons in the cluster to participate in the external flow if they are loosely bound.
A more realistic picture would probably be intermediate between these latter two scenarios, whereby a fraction of neutrons in the cluster are able to `freely' move with the external fluid (see also Fig.~7 in \citet{Martin2016}).

Let us use $\rho_b^{\rm f}$ ($n^{\rm f}$) to denote the mass density (number density) of the neutrons that participate in the flow (therefore $\rho_{b,i} - \rho_b^{{\rm f}}$ is the mass density of neutrons plus protons moving together in the cluster).
The condition of the continuity of the current at the interface between the ion and the external fluid implies the following boundary condition to the velocity potential:
\begin{equation}
  \rho_b^{{\rm f}} \left. \left( \frac{\partial\Phi_{\rm in}}{\partial r}- v_r\right ) \right|_{r=r_N} = 
  \rho_{b,{\rm g}} \left. \left( \frac{\partial\Phi_{\rm out}}{\partial r}- v_r\right ) \right|_{r=r_N} \ .
\end{equation}
This means that we can consider only the $l=1$ component in Eq.~(\ref{eq:poisson-sol}) for $\Phi_{\rm in}$. 
Additionally imposing the continuity of $\Phi$ at $r=r_N$, that the medium is at rest at infinity (i.e. $\Phi_{\rm out} \rightarrow 0$ for $r \rightarrow \infty$), and the non-divergence of $\Phi_{\rm in}$ at $r=0$, the solution of the Laplace equation, Eq.~(\ref{eq:poisson-sol-l1}), reads:
\begin{equation}
\Phi(r,\theta) = 
\begin{cases}
    \frac{\delta^{\rm f}-\gamma}{\delta^{\rm f} + 2\gamma} \ r \ v_r & \text{if } r<r_N \ , \\
    \frac{\delta^{\rm f}-\gamma}{\delta^{\rm f} + 2\gamma} \ \frac{r_N^3}{r^2} \ v_r \ & \text{if } r>r_N \ ,
\end{cases}
\label{eq:phi_out}
\end{equation}
where we define $\delta^{\rm f} \equiv \rho_{b}^{{\rm f}}/\rho_{b,i}$ and $\gamma \equiv \rho_{b,{\rm g}}/\rho_{b,i}$.

The kinetic energy associated to the flow is calculated using Eq.~(\ref{eq:ek-int}), yielding:
\begin{equation}
E_k(\vec{v}) = \frac{1}{2} M_i v_0^2 \left[ 1 - \delta^{\rm f} + \frac{(\delta^{\rm f}-\gamma)^2}{\delta^{\rm f} + 2\gamma} \right] = \frac{1}{2} M_i^\star v_0^2\ ,
\label{eq:ek-sphere}
\end{equation}
where we define the effective mass of the ion as:
\begin{equation}
    M_i^{\star} = M_i \left[ 1 - \delta^{\rm f} + \frac{(\delta^{\rm f}-\gamma)^2}{\delta^{\rm f} + 2\gamma} \right] \ .
\label{eq:m*-sphere}
\end{equation}
As $\delta^{\rm f} > 0$, we can see that the effective mass of the cluster is reduced with respect to the bare mass; in other words, the cluster moves in the medium with a reduced speed.
In the limiting $\delta^{\rm f} = 0$ case, corresponding to the situation where no neutrons participate in the external flow and all neutrons move with the protons in the cluster, one recovers Eq.~(\ref{eq:m*-hard-sphere}).

Again assuming that the neutron and proton mass are equal and that the dripped-proton density is negligible leads to $\delta^{\rm f} \approx n^{\rm f}/n_i$ and $\gamma \approx \nng / n_i$.
Neglecting the possible contribution from proton unbound states appears to be a reasonable approximation at the relatively low temperatures considered in this work (see lower panels of Fig.~\ref{fig:bsk24_betaequi_modifiedftrans_densities_deltaf-eqgamma-linearscalex}).

In this approximation, and setting $\delta^{\rm f} = 1$, meaning that all neutrons in the cluster participate in the `free' motion and the cluster is completely permeable to the flow of the outside fluid, one retrieves Eq.~(12) of \citet{Magierski2004b}, which has the correct high-density physical limit, namely $M_i^{\star} \rightarrow 0$ for $\gamma \rightarrow 1$ (uniform homogeneous system), 
\begin{equation}
    M_i^\star = M_i \frac{(1-\gamma)^2}{1+2\gamma} \ .
\label{eq:m*-mb}
\end{equation}

To precisely evaluate the effective mass, fully microscopic calculations of the dynamical properties of nuclei in a dense nuclear-matter medium would be needed.
Nevertheless, we can evaluate the influence of the effective mass in two additional limiting cases: 
\begin{enumerate}
\item The first considers that only neutrons in the continuum participate in the (collective) flow, as proposed by \citet{Carter}.
In this case, we can write $\delta^{\rm f}_{\rm min} = \rho_{b,{\rm unbound}}/\rho_{b,i} \approx \nng / n_i$, leading to 
\begin{eqnarray}
    M_i^\star &=& M_i \left[ 1 - \delta^{\rm f}_{\rm min} + \frac{(\delta^{\rm f}_{\rm min}-\gamma)^2}{\delta^{\rm f}_{\rm min} + 2\gamma} \right] \nonumber \\
     & \approx & M_i \left(1- \frac{\nng}{n_i} \right) \ ,
    \label{eq:m*-deltamin}
\end{eqnarray}
where we assume that the continuum contribution is homogeneous.
It is interesting to note that here the effective mass is equal to the nucleus mass in the e-cluster representation proposed by \citet{papa2013} and \citet{Raduta2014}. 
Although the prescription for continuum counting as $\rho_{b,{\rm unbound}}/\rho_{b,i} \approx \nng / n_i$ is rough, it is found to give an accurate estimation of the total number of  unbound non-resonant states obtained from a microscopic HF \citep{papa2013} or HFB  \citep{Chamel2012,Martin2016} calculation in the whole density region corresponding to the inner crust. 
\item In the second limiting case, all neutrons in the cluster are considered to participate in the (external) flow, that is, $\delta_{\rm max} = \rho_{b,{\rm i}n}/\rho_{b,i} \approx 1 - Z/A \equiv 1 - y_p$, leading to 
\begin{eqnarray}
    M_i^\star &=& M_i \left[ 1 - \delta^{\rm f}_{\rm max} + \frac{(\delta^{\rm f}_{\rm max}-\gamma)^2}{\delta^{\rm f}_{\rm max} + 2\gamma} \right] \nonumber \\
    &\approx& M_i \left[ y_p + \frac{(1-y_p-\nng/n_i)^2}{1-y_p + 2 \nng/n_i} \right] \ .
    \label{eq:m*-deltamax}
\end{eqnarray}
\end{enumerate}

The use of the different expressions for the ion effective mass in the translational free energy is presented in Sect.~\ref{sec:trans}.

\section{Model of the inner crust at finite temperature}
\label{sec:formalism}
In the OCP or single-nucleus approximation, the inner crust of a NS can be considered to be composed of identical WS cells, each of which includes a fully ionised ion characterised by mass number $A$ and proton number $Z$, radius $r_N$, and internal density $n_i$.
In the solid phase, the ion is located at the centre of the (body-centred cubic; bcc) cell of volume $V_{\rm WS}$ and it is surrounded by uniform distributions of electrons, neutrons, and protons, whose densities are respectively denoted by $n_e$, $n_{{\rm g}n}$, and $n_{{\rm g}p}$.
In the liquid phase we are interested in, because of the OCP approximation, a WS volume can still be defined as the optimal volume centred on each (moving) ion, and is obtained from the condition of charge neutrality; see Eq.~(\ref{eq:charge}).

At a given thermodynamic condition, defined by the baryonic density $n_B$ and temperature $T$ if beta equilibrium is assumed\footnote{In the case where the total proton fraction $Y_p^{\rm tot}$ is fixed, the thermodynamic condition is defined by the baryonic density $n_B$, the temperature $T$, and $Y_p^{\rm tot}=n_e/n_B$.}, the total free-energy density of the system can be written as
 \begin{equation}
     \mathcal{F} = \mathcal{F}_e +  \mathcal{F}_{\rm g} (1-u) + \frac{F_i}{V_{\rm WS}}, 
     \label{eq:total_free_energy_density}
 \end{equation}
where $\mathcal{F}_e =\mathcal{F}_e(n_e, T)$ is the electron gas free-energy density\footnote{We use the uppercase $F$, the lowercase $f$, and $\mathcal{F}$ to denote the free energy per cell, the free energy per nucleon, and the free energy per unit volume, respectively.}, 
$\mathcal{F}_{\rm g} =\mathcal{F}_B(n_{\rm g}, \delta_{\rm g}, T) + m_n c^2 n_{{\rm g}n} + m_p c^2 n_{{\rm g}p}$ is the free-energy density of uniform nuclear matter (including the rest masses of nucleons) at baryonic density $n_{\rm g} = n_{{\rm g}n} + n_{{\rm g}p}$, isospin asymmetry $\delta_{\rm g} =\frac{n_{{\rm g} n} - n_{{\rm g}p}}{n_{{\rm g}n} + n_{{\rm g}p}} $, and temperature $T$ (see Sect.~\ref{sec:nucfunc} for more details),
$u=\frac{A/n_i}{V_{\rm WS}}$ 
is the volume fraction of the cluster, and $F_i$ is the cluster free energy, which includes the Coulomb interaction between the nucleus and the electron gas, as well as the residual interface interaction between the nucleus and the surrounding dilute nuclear-matter medium (see Sect.~\ref{sec:finsize} for details).
The excluded-volume term breaking the additivity of the free energy in Eq.~(\ref{eq:total_free_energy_density}), $-u\mathcal{F}_{\rm g}$, has also been shown to account for the subtraction of the gas states from the nuclear partition sum, which avoids double counting of unbound single-particle states \citep{gulrad2015,Mallik2021}.
 
The total free-energy density $\mathcal{F}$ in Eq.~(\ref{eq:total_free_energy_density}) was minimised under the two constraints coming from the conservation of the baryonic number and the charge neutrality:
 \begin{align}
     n_B &= \frac{2n_p}{(1-I)}\left(1 - \frac{n_{{\rm g}n} + n_{{\rm g}p}}{n_i}\right) + n_{{\rm g}n} + n_{{\rm g}p},\label{eq:baryon_conservation}\\
     n_e &= n_p + n_{{\rm g}p}\left(1- \frac{2n_p}{n_i (1-I)}\right),
        \label{eq:charge}
 \end{align}
where $I = 1-2Z/A$ and $n_p = Z/V_{\rm WS}$. 
Two Lagrange multipliers, $\gamma_1$ and $\gamma_2$, were introduced, and the function to be minimised is expressed as 
\begin{equation}
                \begin{split}
                        \Omega&= \frac{2n_p}{1-I}\frac{F_i}{A}+ \left[ 1 - \frac{2n_p}{(1-I)n_i}\right]\mathcal{F}_{\rm g} + \mathcal{F}_e \\
                        & + \gamma_1 \left[n_B - \frac{2n_p}{1-I}\left(1 - \frac{n_{{\rm g}n} + n_{{\rm g}p}}{n_i}\right) - n_{{\rm g}n} -n_{{\rm g}p}\right] \\
                        &+ \gamma_2 \left[n_e - n_p - n_{{\rm g}p}\left(1- \frac{2n_p}{n_i (1-I)}\right)\right].
        \end{split}
        \label{eq:Omega}
\end{equation}
Minimising $\Omega$ with respect to  $n_{{\rm g}n}$ and $n_{{\rm g}p}$, we obtain the following expressions for $\gamma_1$ and $\gamma_2$:
\begin{align}
                        \gamma_1 &= \mu_{n} \label{eq:gamma_1}\\
            \gamma_2 &= -\mu_n + \mu_{p} 
            \label{eq:gamma_2},
\end{align}
in which $\mu_n$ and $\mu_p$ are respectively defined as
\begin{align}
\mu_n & \equiv \frac{\partial \mathcal{F}_{\rm g}}{\partial n_{{\rm g}n}}  + \frac{2n_pn_i}{n_i(1-I)- 2n_p} \frac{\partial(F_i/A)}{\partial n_{{\rm g}n}}, \label{eq:mu_n} \\
\mu_p & \equiv \frac{\partial \mathcal{F}_{\rm g}}{\partial n_{{\rm g}p}}  + \frac{2n_pn_i}{n_i(1-I)- 2n_p} \frac{\partial(F_i/A)}{\partial n_{{\rm g}p}}. \label{eq:mu_p}
\end{align}
On the right-hand side of Eqs.~(\ref{eq:mu_n})-(\ref{eq:mu_p}), we can easily see that the first terms are the chemical potentials of the unbound nucleons, $\mu_{{\rm g}n ({\rm g}p)} \equiv \left.\frac{\partial \mathcal{F_{\rm g}}}{\partial n_{{\rm g}n ({\rm g}p)}}\right|_{n_{{\rm g}p ({\rm g}n) }}$, while the last terms account for the in-medium modification induced by the Coulomb screening and the centre-of-mass translation.

In order to describe the nuclear energetics, we employed a compressible liquid-drop (CLD) model approach, as in \citet{Carreau2019}, \citet{Carreau2020}, \citet{dinh2021}, and \citet{dinhEPJA21}, which describes the ion as a cluster of nucleons in a leptodermous expansion. 
Considering the inner crust in the liquid phase, the collective degrees of freedom are translational, meaning that $F_i$ in Eq.~(\ref{eq:total_free_energy_density}) can be written as:
\begin{equation}
    F_i = M_i c^2 + F_{\rm bulk} + F_{\rm Coul + surf +  curv} + F_{\rm trans},
    \label{eq:Fi0}
\end{equation}
where $M_i = (A-Z)m_n + Zm_p$ is the total bare mass of the cluster, $F_{\rm bulk} = \frac{A}{n_i} \mathcal{F}_B(n_i, I, T)$ is the cluster bulk free energy, ${F}_{\rm Coul + surf +  curv}$ is the sum of the Coulomb, surface, and curvature energies, and the last term, $F_{\rm trans}$, accounts for the translational degree of freedom of the cluster. 
We discuss these terms in detail in the following sections.

\subsection{Nuclear matter free energy in the mean-field approximation} 
\label{sec:nucfunc}
Computation of the cluster and the surrounding nuclear-matter bulk free-energy terms requires knowledge of the free-energy density of homogeneous nuclear matter $\mathcal{F}_B(n, \delta, T)$ at a total baryonic density\footnote{In this section, $n_p$ refers to the proton density in homogeneous matter, while elsewhere in the paper $n_p=Z/V_{\rm WS}$.} $n=n_n+n_p$ and isospin asymmetry $\delta=(n_n-n_p)/n$.  
To this aim, we used the self-consistent mean-field thermodynamics \citep{lattimer1985,Ducoin2007}. 
In this approach, the free energy was decomposed into a `potential' and a `kinetic' part:
\begin{equation}
   \mathcal{F}_B(n, \delta, T) =  \mathcal{V}(n, \delta) + \mathcal{F}_{\rm kin}(n, \delta, T)  , 
   \label{eq:HMenergy}
\end{equation}
where the temperature dependence encoded in the term $\mathcal{F}_{\rm kin}$ is the same as in a system of independent quasi-particles with effective single-particle energies $e_{{q}}=p^2/2m^{\star}_q$ ($p$ being the momentum and ${q} = n, p$ labelling neutrons and protons) and with shifted effective chemical potentials $\tilde{\mu}_q$ that account for the self-consistent interaction. 

The terms `potential' and `kinetic' are put into quotes, because the kinetic term includes density-dependent neutron and proton effective masses $m^{\star}_q(n_p,n_n)$ that physically arise from the non-locality of the effective nucleon--nucleon interaction. 
These effective masses lead to a deviation from the parabolic approximation to the symmetry energy \citep{Burgio2021,Grams2022}, and a complex temperature dependence of the bulk term as discussed below, both for the dripped nucleons and for the nucleus.
Beyond-mean-field effects may lead to an extra explicit temperature dependence of the potential term $\mathcal{V}$. However, effects beyond the renormalisation of the nucleon effective mass were shown to be very small both in Brueckner-Hartree-Fock \citep{Burgio2019,Burgio2020} and in many-body perturbation theory \citep{Somasundaram2021}, and are neglected in this work. 

The density dependence of the `potential' term was set to reproduce different nuclear functionals at zero temperature, covering the uncertainties of nuclear theory.
To this aim, we employed the so-called meta-modelling approach proposed in \citet{Margueron2018a, Margueron2018b}, where $\mathcal{V}=\mathcal{V}^{N}_{\rm MM}$ was expressed as a Taylor expansion truncated at order $N$ around the saturation point $(n=n_{\rm sat},\delta=0)$. 
 \cite{Margueron2018a}  showed that different nucleonic energy functionals at zero temperature can be satisfactorily reproduced by truncating the expansion at order $N=4$:
\begin{equation}
   \mathcal{V}^{N = 4}_{\rm MM}(n, \delta) = \sum_{k=0}^{4} \frac{n}{k!}(v^{\rm is}_{k} +v^{\rm iv}_{k}\delta^2 )x^{k}u^{N = 4}_{k}(x),
   \label{eq:vMM}
\end{equation}
where $x = (n - n_{\rm sat})/(3n_{\rm sat})$ and $u^{N}_{k}(x) = 1 - (-3x)^{N+1-k}\exp(-b(1+3x))$, with $b = 10\ln(2)$ being a parameter governing the function at low densities. 
The factor $u^{N}_{k}(x)$ was introduced to ensure the convergence at the zero-density limit, while the parameters $v^{\rm is}_{k}$ and $v^{\rm iv}_{k}$ are linear combinations of the so-called nuclear matter empirical parameters, $\{ n_{\rm sat},E_{\rm sat,sym},L_{\rm sym},K_{\rm sat,sym},Q_{\rm sat,sym},Z_{\rm sat,sym}\}$ (see \citet{Margueron2018a} for details). 
In this work, we mainly employ the BSk24 functional \citep{BSK24} as an illustrative example, but we also consider the SLy4 \citep{SLy4} and DD-ME$\delta$ \citep{DDMEd} functionals to study the model dependence of the results (see Sect.~\ref{sec:result_betaequi}). 
The numerical values of the nuclear matter empirical parameters of these models, including those ruling the density dependence of the nucleon effective masses, are listed in Table~1 of \citet{dinh2021}. 
These three interactions were selected because they all fulfill the basic constraints from nuclear theory, nuclear experiments, and NS observables, and at the same time they reasonably cover present uncertainties on the nucleonic energy functionals~\citep{dinh2021, DinhThi_Universe}.

To evaluate the $\mathcal{F}_{\rm kin}(n, \delta, T)$ term,  we used the thermodynamic relation 
 \begin{equation}
 \mathcal{F}_{\rm kin} =\sum_{q=n,p} \mathcal{E}^{q}_{\rm kin} - T \mathcal{S}^{q}_{\rm kin}, \label{eq:Fkin}
 \end{equation}
where $\mathcal{E}^{q}_{\rm kin}$ and $\mathcal{S}^{q}_{\rm kin}$ are the kinetic energy and entropy densities, respectively, of a system of independent particles with  single-particle energy $e_{{q}}=p^2/2m^{\star}_q $. The kinetic energy density corresponding to particles of type $q$ at density $n_q$ (the density of the other species being noted as $n_{q^\prime}$) is given by:
\begin{equation}
    \mathcal{E}^{q}_{\rm kin} =  \int_0^{\infty} de_{q} \frac{e_{q} \rho(e_{q})}{1 + \exp{\frac{e_{q} - \Tilde{\mu}_{{q}}}{k_{\rm B} T}}}
    =\frac{3 k_{\rm B} T}{\lambda_q^3}F_{3/2}\left(\frac{\Tilde{\mu}_{{q}}}{k_{\rm B} T}\right),
    \label{eq:Ekin}
\end{equation}
where $\rho(e_{{q}})$ is the density of fermionic energy states,
\begin{equation}
\rho(e_{{q}})= \frac{1}{2\pi^2}\left(\frac{2m_{{q}}^{\star}}{\hbar^2}\right)^{3/2}\sqrt{e_{{q}}} \ ,
\end{equation}
$\hbar$ is the Planck constant, $m_{{q}}^{\star}(n_q,n_{q^\prime})$  is the density-dependent nucleon effective mass, 
$\lambda_{q} = \left(\frac{2\pi \hbar^2}{k_{\rm B} T m^{\star}_{q}}\right)^{1/2}$ is the nucleon thermal wavelength, and $F_{3/2}$ denotes the Fermi-Dirac integral:
\begin{equation}
    F_j(x) = \frac{1}{\Gamma(j+1)}\int_{0}^{\infty}\frac{t^jdt}{1+ e^{t-x}},
    \label{eq:FD_integral}
\end{equation}
where $j > -1$.
The system of equations is closed by the relation between the effective chemical potential $\Tilde{\mu}_{q}$ and the particle density $n_{q}$:
\begin{equation}
    n_{{q}} = \int_0^{\infty} de_{q} \frac{\rho(e_{q})}{1 + \exp{\frac{e_{q} - \Tilde{\mu}_{{q}}}{k_{\rm B} T}}}=\frac{2}{\lambda_{{q}}^3}F_{1/2}\left(\frac{\Tilde{\mu}_{{q}}}{k_{\rm B} T}\right). \label{eq:n_q}
\end{equation}
The entropy is defined from the partial grand canonical thermodynamic potential $\Phi^{{q}}_{\rm kin}$ as:
\begin{equation}
    \mathcal{S}^{{q}}_{\rm kin} = -\left. \frac{\partial \Phi^{{q}}_{\rm kin}}{\partial T}\right|_{\Tilde{\mu}_{{q}}}. \label{eq:entropy}
\end{equation}
Using the expression of $\Phi^{{q}}_{\rm kin}$,
\begin{eqnarray}
    \Phi^{{q}}_{\rm kin} &=& -k_{\rm B} T\int_0^{\infty}
    \ln\left [1 + \exp{\left(-\frac{e_{{q}} - \Tilde{\mu}_{{q}}}{k_B T}\right)}\right ]\rho(e_{{q}})de_{{q}}\nonumber \\ &=&-\frac{2}{3}\mathcal{E}^{q}_{\rm kin} \ ,
    \label{eq:Phi_q}
\end{eqnarray}
we obtain:
\begin{equation}
    \mathcal{S}^{{q}}_{\rm kin} = \frac{5}{3}\frac{\mathcal{E}^{q}_{\rm kin}}{T}-\frac{\Tilde{\mu}_q}{T}n_q. \label{eq:entropy2}
\end{equation}
Incorporating Eqs.~(\ref{eq:entropy2}) and (\ref{eq:Ekin}) into Eq.~(\ref{eq:Fkin}), the final result for $\mathcal{F}_{\rm kin}$ reads:
\begin{equation}
    \mathcal{F}_{\rm kin} = \sum_{q = n,p} \left[ { \frac{-2 k_{\rm B} T}{\lambda_{q}^3}F_{3/2}\left(\frac{\Tilde{\mu}_{{q}}}{k_{\rm B} T}\right) + n_q \Tilde{\mu}_{{q}}} \right].
   \label{eq:Fkin2}
\end{equation}

Finite-temperature mean-field theory \citep{lattimer1985, Ducoin2007} implies that the free-energy density $\mathcal{F}_{\rm kin}$ given by Eq.~(\ref{eq:Fkin2}) represents the correct `kinetic' term entering  Eq.~(\ref{eq:HMenergy}), provided the following relation is imposed between the effective chemical potential $\Tilde{\mu}_{q}$ and the thermodynamic potential $\mu_q$ in Eqs.~(\ref{eq:mu_n})-(\ref{eq:mu_p}):
\begin{equation}
\Tilde{\mu}_{q}=\mu_q-U_q,
\end{equation}
where the mean-field potential $U_q$ is given by:
\begin{equation}
U_q = \left. \frac{\partial \mathcal{V}^{N}_{\rm MM}}{\partial n_q}\right|_{n_{q^\prime}} + \sum_{\bar{q} = n,p}\mathcal{E}^{\bar{q}}_{\rm kin} 
m^{\star}_{\bar{q}} \left. \frac{\partial \left(1/ m^{\star}_{\bar{q}}\right)}{\partial n_q} \right|_{n_{q^\prime}} \ ,
\end{equation}
and the partial derivatives with respect to $n_n$ ($n_p$) are computed keeping $n_p$ ($n_n$) constant.

\subsection{Coulomb, surface, and curvature energies}
\label{sec:finsize}
To model the inhomogeneities in the inner crust, in Eq.~(\ref{eq:Fi0}) the bulk free energy $F_{\rm bulk}$ was complemented by Coulomb, surface, and curvature contributions: 
\begin{equation}
F_{\rm Coul + surf +  curv} = V_{\rm WS}(\mathcal{F}_{\rm Coul} + \mathcal{F}_{\rm surf} + \mathcal{F}_{\rm curv})\ .
\end{equation}
For the latter terms, we used the same expressions as in \citet{Carreau2019}, \citet{Carreau2020}, \citet{dinh2021}, and \citet{dinhEPJA21}, which are reported here for completeness.
The Coulomb term giving the electrostatic proton--proton, proton--electron, and electron--electron interaction energies, assuming a spherical geometry for the WS cell, reads:
\begin{equation}
    \mathcal{F}_{\rm Coul}  = 2\pi (en_ir_N)^2\left(\frac{1-I}{2} -\frac{n_{{\rm g}p}}{n_i}\right)^2 u \eta_{\rm Coul}(u), \label{eq:Fcoul}
\end{equation}
where $e$ is the elementary charge, and the function $\eta_{\rm Coul}(u)$ accounting for the electron screening is written as \citep{Ravenhall1983_prl, Pethick1995}
\begin{equation}
    \eta_{\rm Coul}(u) =\frac{1}{5}\left [ u+ 2 \left ( 1- \frac{3}{2}u^{1/3} \right ) \right ]. \label{eq:eta_Coul}
\end{equation}
Regarding the surface and curvature energies, we employ the expression of \citet{lattimer1991}, \citet{Maru2005}, and \citet{Newton2013}:
\begin{equation}
\mathcal{F}_{\rm {surf}} + \mathcal{F}_{\rm {curv}} =\frac{u d}{r_N}\left ( \sigma_{\rm s}(I, T) +\frac{(d-1)\sigma_{\rm c}(I, T)}{r_N}\right ) , \label{eq:interface}   
\end{equation}
where $d=3$, because here we only consider spherical nuclei, and $\sigma_{\rm s}$ and $\sigma_{\rm c}$ are the surface and curvature tensions. 
At zero temperature, we adopt the expressions of $\sigma_{\rm s} (I, T = 0)$ and $\sigma_{\rm c}(I, T=0)$ from \citet{Ravenhall1983}, which are based on Thomas-Fermi calculations at extreme isospin asymmetries:
\begin{eqnarray}
\sigma_{\rm s}(I, T = 0) &=& \sigma_0\frac{2^{p+1}+b_s}{y_p^{-p}+b_s+(1-y_p)^{-p}} \ , \label{eq:surface} \\
\sigma_{\rm c}(I, T = 0)&=&5.5 \, \sigma_{\rm s}(I, T = 0) \frac{\sigma_{0, {\rm c}}}{\sigma_0}(\beta-y_p)\ , \label{eq:curvature} \ 
\end{eqnarray}
where $y_p = (1-I)/2$, and the parameters $(\sigma_0,\sigma_{0,{\rm c}},b_s,\beta,p)$ are optimised for each given bulk and effective mass functional in order to reproduce the experimental nuclear masses in the atomic mass evaluation (AME) 2016 \citep{AME2016}.

At finite temperature, the surface tension is modified with respect to the case of zero temperature, and we use the following expression \citep{lattimer1991}:
\begin{equation}
\sigma_{\rm s,c}(I, T) = \sigma_{\rm s,c}(I, T =0)h(T),
                \label{eq:sigmas_T}
\end{equation}
where
\begin{equation}
h(T) = 
\begin{cases} 
  0 & \text{if $T > T_c$} \ ,  \\
  \left[1 -\left(\frac{T}{T_c}\right)^2\right]^2 & \text{if $T \leq T_c$} \ ,
 \end{cases}
\end{equation}
and $T_c$ is the critical temperature given by Eq.~(2.31) in \citet{lattimer1991}.

\subsection{Translational energy}
\label{sec:trans}
As the temperatures of interest in this work are  above the melting point, clusters are expected to be in the liquid phase. As a result, it is necessary to account for their translational degrees of freedom. 

\begin{figure}
    \centering
    \includegraphics[scale = 0.45]{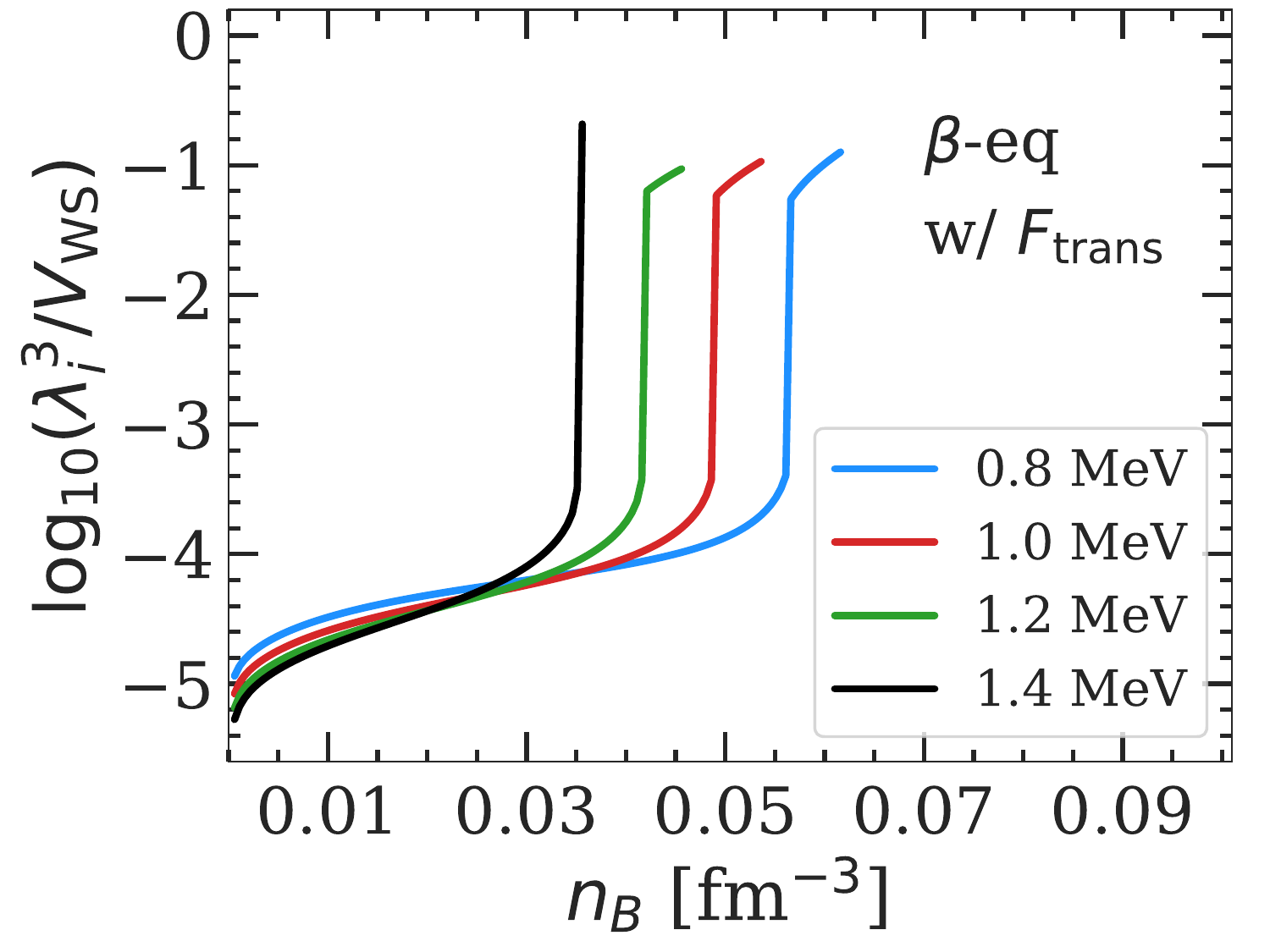}
    \caption{Ratio (in logarithmic scale) of the ion thermal wavelength to the WS cell volume as a function of the baryonic density $n_B$ for the BSk24 model at four different temperatures: $k_{\rm B} T = 0.8$~MeV (blue line), $k_{\rm B} T = 1.0 $~MeV (red line), $k_{\rm B} T = 1.2$~MeV (green line), and $k_{\rm B} T = 1.4$~MeV (black line). The results are obtained at beta equilibrium using the ideal-gas translational energy, Eq.~(\ref{eq:ftrans}).}
\label{fig:ratio-lambda-Vws}
\end{figure}

If the effect of the fluid of dripped nucleons is neglected, the free energy corresponding to the translational motion reduces to the standard expression of an ideal gas \citep{hpy2007}:
\begin{equation}
    F_{\rm trans}= k_{\rm B} T\ln \left( \frac{1}{V_{\rm WS}}\frac{\lambda_i^{3}}{g_s}\right)  - k_{\rm B} T. 
    \label{eq:ftrans}
\end{equation}
This expression is easily obtained considering a system of $N_{\rm tot}$ identical ions in a volume  $V_{\rm tot}$. In the thermodynamic limit, $N_{\rm tot} \rightarrow \infty$, $V_{\rm tot} \rightarrow \infty$, and $N_{\rm tot}/V_{\rm tot} \rightarrow 1/V_{\rm {WS}}$. 
The partition function for the centre-of-mass motion of a non-relativistic classical particle can be written as
\begin{equation}
    Z^{\rm cl}_1 = g_s V_{\rm tot}\left (\frac{M_i}{h}\right)^3\int\exp{\left(-\frac{M_i v^2}{2 k_{\rm B} T}\right)}d^3v = \frac{V_{\rm tot}g_s}{\lambda_i^3} \ ,
 \label{eq:zcl}
\end{equation}
where $\lambda_i = \sqrt{\frac{2\pi \hbar^2}{M_i k_{\rm B} T}}$ is the ion thermal wavelength, and $g_s=2J_i+1$ is the ground-state spin degeneracy, which we set to unity independent of the ion species, because  $J_i=0$ for even-even nuclei, which are most abundant in the crust. 
This expression is valid in the non-degenerate Boltzmann limit, corresponding to $\lambda_i^3 \ll  V_{\rm WS}$.
For a system of $N_{\rm tot}$ identical and indistinguishable clusters, the total partition sum reads:
\begin{equation}
    Z^{\rm cl}_{\rm tot} = \frac{(Z^{\rm cl}_1)^{N_{\rm tot}}}{N_{\rm tot}!}.
    \label{eq:Zcl_tot}
\end{equation}
Using $F_{\rm trans} = {-k_{\rm B} T \ln{Z^{\rm cl}_{\rm tot}}}/{N_{\rm tot}}$, Eq.~(\ref{eq:ftrans}) is immediately found.
To check the validity of the non-degenerate limit for the calculation of the translational free energy, Eq.~(\ref{eq:ftrans}), we plot in Fig.~\ref{fig:ratio-lambda-Vws} the ratio of the ion thermal wavelength, $\lambda_i^3$, to the WS cell volume as a function of the inner-crust baryon number density for the BSk24 model in beta equilibrium and for four different temperatures.
We can see that, although the ratio $\lambda_i^3/V_{\rm WS}$ remains smaller than $1$ throughout the inner crust, the non-degenerate Boltzmann limit tends to break at high densities and higher temperatures in the vicinity of the crust--core transition.

The derivation of Eq.~(\ref{eq:ftrans}) also shows that this expression is only valid in the limit of heavy ions, that is, those with $M_i/m_n \gg 1$, and if ions are considered as point-like particles moving in vacuum.
The first of these conditions is well realised in the inner crust, but the latter two clearly become more doubtful as density increases. 
A first obvious correction to Eq.~(\ref{eq:ftrans}) that was considered in the literature since the first works on the finite-temperature crust \citep{lattimer1985, lattimer1991} is to consider a reduced volume $V_{\rm f}$ for the centre-of-mass motion, in coherence with the excluded-volume approach of Eq.~(\ref{eq:total_free_energy_density}):
\begin{equation}
   V_{\rm WS} \rightarrow V_{\rm f} = \frac{4}{3} \pi (r_{\rm WS} - r_{N})^3 \ ,
\end{equation}
with $r_{\rm WS}$ being the WS cell radius.

Moreover, in a proto-NS inner crust, nuclei are surrounded by a distribution of dripped protons and neutrons. 
At the bottom of the inner crust, the densities of these fluids are comparable to the nuclear internal density and, as a result, the cluster bare mass $M_i$ should be replaced by an effective mass $M_i^{\star}$, accounting for the modification of the free flow in a nuclear medium.

In Sect.~\ref{sec:hydro} we present the derivation of the effective mass of the cluster moving in a uniform nucleon medium modelled as an ideal fluid, with different hypotheses as to the fluid properties of the cluster. 
Repeating the calculation of the translational partition sum of Eq.~(\ref{eq:zcl}) ---using either Eq.~(\ref{eq:ek-hard-sphere}) or Eq.~(\ref{eq:ek-sphere}) for the thermal kinetic energy--- leads to an expression identical to Eq.~(\ref{eq:ftrans}), with the substitution $M_i\rightarrow M_i^{\star}$.
Including both the corrections on the finite-size effect of nuclei and the effective mass, the translational free energy can be written as
\begin{equation}
    {F}_{\rm trans}^\star= k_{\rm B} T\ln \left( \frac{1}{V_{\rm f}}\frac{\lambda_i^{\star 3}}{g_s}\right)  - k_{\rm B} T, 
    \label{eq:ftrans_eff}
\end{equation}
where $\lambda^{\star}_i = \sqrt{\frac{2\pi \hbar^2}{M^{\star}_i k_{\rm B} T}}$,
with $M_i^\star$ given by Eq.~(\ref{eq:m*-sphere}).
For the latter term, we considered three cases:
\begin{enumerate}
\item the limiting case $\delta^{\rm f}=0$, leading to Eq.~(\ref{eq:m*-hard-sphere}), corresponding to a solid cluster fully impermeable to the surrounding nuclear medium;
\item the limiting case $\delta^{\rm f}=1$, corresponding to the single-fluid calculation by \citet{Magierski2004b}, Eq.~(\ref{eq:m*-mb});
\item the intermediate case $\delta^{\rm f}= \delta^{\rm f}_{\rm min} \approx \gamma$, leading to Eq.~(\ref{eq:m*-deltamin}).
We think that this latter case, which considers that the translational motion concerns protons, bound neutrons, and neutrons in unbound resonant states, while all neutrons occupying continuum states constitute the external fluid, is more realistic. 
Indeed, the mobility of the nucleons is given by their velocity flux at the Fermi surface \citep{Carter}. 
As such, this is only defined for `conduction' nucleons with single-particle wave functions, which do not vanish at the WS boundary, that is, continuum states as defined in \citet{papa2013}. 
Also,  \cite{gulrad2015} showed that the factorization of the partition sum that is needed to theoretically obtain the extended nuclear statistical equilibrium models used to produce the general-purpose equation of state \citep{Oertel2017, Compose,Raduta2019} is only possible if the cluster mass appearing in the translational part is the e-cluster mass, $A_e=A(1-n_{\rm g}/n_i)$. 
\end{enumerate}

\begin{figure}
    \centering
   \includegraphics[scale = 0.38]{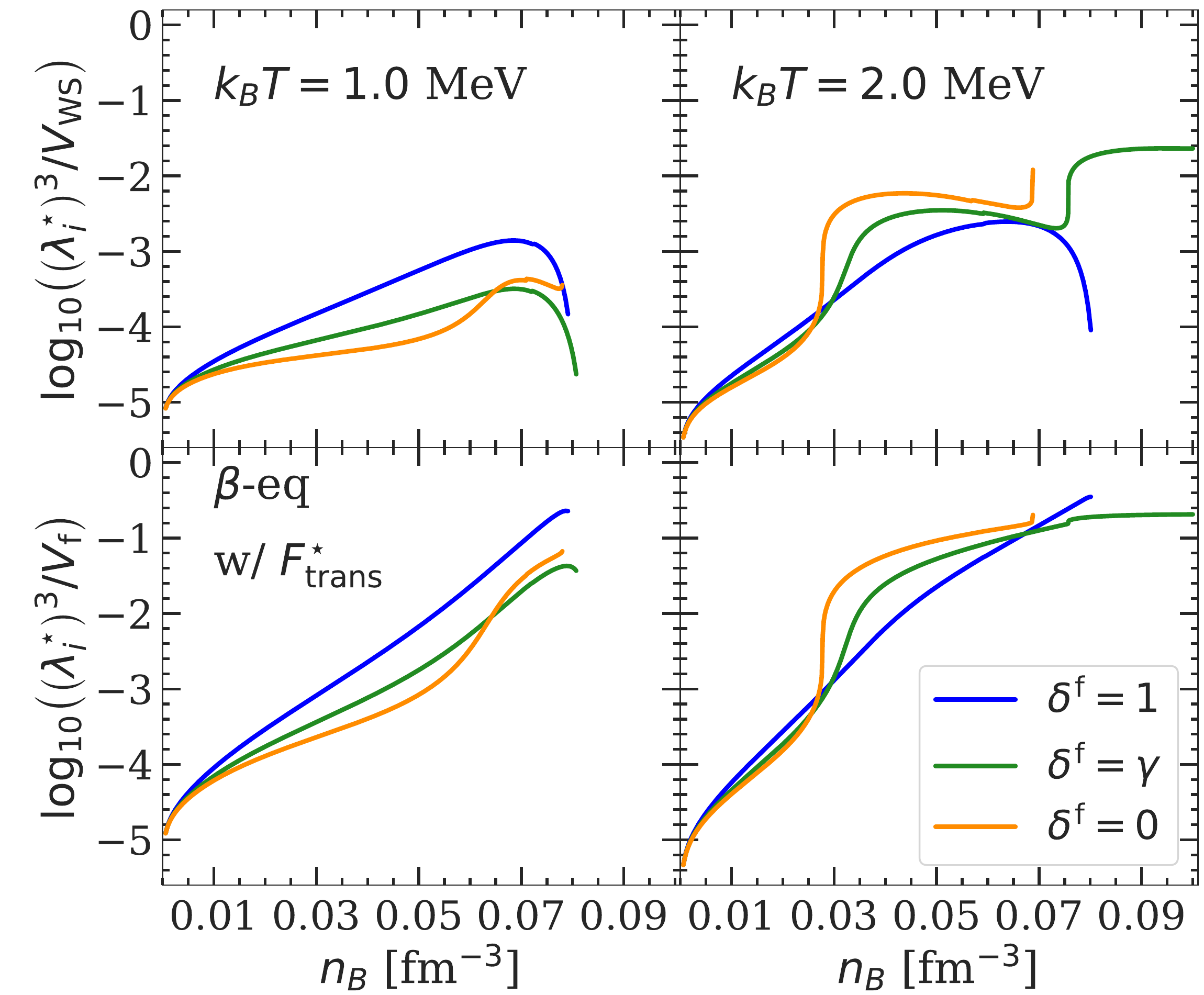}
    \caption{Ratio (in logarithmic scale) of the ion thermal wavelength to the WS cell volume (top panels) and to the reduced volume (bottom panels) as a function of the baryonic density $n_B$ for the BSk24 model, using three different prescriptions for the ion effective mass, at two different temperatures: $k_{\rm B} T = 1$~MeV (left panels) and $k_{\rm B} T = 2.0 $~MeV (right panels). The results are obtained at beta equilibrium using the translational energy $F^\star_{\rm trans}$, Eq.~(\ref{eq:ftrans_eff}), accounting for the in-medium correction.}
\label{fig:ratio-lambda-Vws-eff}
\end{figure}

For comparison with Fig.~\ref{fig:ratio-lambda-Vws},  Fig.~\ref{fig:ratio-lambda-Vws-eff} shows the ratio of the ion thermal wavelength, $(\lambda_i^\star)^3$, to the WS cell volume $V_{\rm WS}$ (top panels) as a function of the inner-crust baryonic density $n_B$ for the BSk24 model at beta equilibrium and at two different temperatures: $1.0$~MeV (left panels) and $2.0$~MeV (right panels).
Three different prescriptions of $M_i ^\star$  are considered, corresponding to the three listed cases of $\delta^{\rm f}$.
We observe that, in this case, the ratio $(\lambda_i^\star)^3/V_{\rm WS}$ remains always $\ll 1$ and smaller than that obtained in the ideal-gas case near the crust--core transition.
This is due to the fact that in the ideal-gas case (taking $M_i \approx A m_n$), when the clusters dissolve near the crust--core transition point, $A$ and $V_{\rm WS}=Z/n_p$ become small (because of the sharp decrease in the cluster mass and proton number; see also Fig.~\ref{fig:betaqui_AZvsnb_bsk24} and related discussion), and therefore the ratio $(\lambda_i)^3/V_{\rm WS}$ is relatively large.
On the other hand, when $F_{\rm trans}^\star$ is used, the clusters remain large until high $n_B$ (see also Figs.~\ref{fig:bsk24_betaequi_modifiedftrans_AZ_3meff-linearcalex} and \ref{fig:betaequi_modifiedftrans_AZ}), yielding a greater cluster mass and WS volume, resulting in a smaller thermal-wavelength-to-WS-cell-volume ratio. 
However, the ratio of $(\lambda_i^\star)^3$ to the reduced volume $V_{\rm f}$ (bottom panels in Fig.~\ref{fig:ratio-lambda-Vws-eff}) increases towards the crust--core transition, reaching $\approx 0.5$ almost independently of the temperature and effective-mass estimation. This shows that the non-degenerate limit implicit in Eq.~(\ref{eq:ftrans_eff}) is always a reasonable approximation, albeit less justified at the highest densities.  

To investigate the effect of the translational free energy on the (proto-)NS inner-crust properties, we performed our calculations employing both the ideal-gas expression (see Eq.~(\ref{eq:ftrans})), and the expression accounting for in-medium effects (see Eq.~(\ref{eq:ftrans_eff})).
The corresponding results are discussed in the following sections.

\section{Results at beta equilibrium}
\label{sec:result_betaequi}

In this section, we discuss the impact of the translational free energy on the properties of the inner crust, assuming that beta equilibrium holds, which is believed to be the case in the proto-NS at moderate temperatures (see \citet{Oertel2017} and references therein). 
In this case, we minimised the function $\Omega$ defined in Eq.~(\ref{eq:Omega}) with respect to the variational variables ($r_N$, $I$, $n_i$, $n_p$, $n_{{\rm g}n}$, $n_{{\rm g}p}$, $n_e$). 
This resulted in the following system of equilibrium equations:
\begin{align}
  \mu_{p} + \mu_e &= \mu_{n}, \label{eqocp:betaequi}\\
         \frac{\partial (F_i/A)}{\partial r_N} &= 0, \label{eqocp:rn}\\
  n_i^2\frac{\partial}{\partial n_i}\left(\frac{F_i}{A}\right) &= P_{\rm g} , \label{eqocp:ni}\\
         \frac{F_i}{A} + (1-I) \frac{\partial}{\partial I}\left(\frac{F_i}{A}\right)&= \mu_n - \frac{P_{\rm g}}{n_i}, \label{eqocp:I}\\
  2\left[  \frac{\partial}{\partial I}\left(\frac{F_i}{A}\right) - \frac{n_p}{1-I}  \frac{\partial}{\partial n_p}\left(\frac{F_i}{A}\right) \right] &=\mu_n - \mu_p, \label{eqocp:Iandnp}
\end{align}
where $P_{\rm g} = \mu_{n} n_{{\rm g}n} +\mu_{p} n_{{\rm g}p} - \mathcal{F}_{\rm g} $ is the pressure of the dripped nucleons, and $\mu_e = \partial \mathcal{F}_e/\partial n_e$ is the electron chemical potential.
It is to be noted that, because of the definition of the chemical potentials $\mu_n$ and $\mu_p$ (see Eqs.~(\ref{eq:mu_n})-(\ref{eq:mu_p})), the pressure term $P_{\rm g}$ is not just the pressure of a self-interacting nucleon gas but also includes in-medium effects.
At each given $(n_B, T)$, this system of five coupled differential equations, Eqs.~(\ref{eqocp:betaequi})-(\ref{eqocp:Iandnp}), was solved numerically together with the two constraints from the baryon number conservation, Eq.~(\ref{eq:baryon_conservation}), and charge neutrality, Eq.~(\ref{eq:charge}), yielding the equilibrium composition.

\begin{figure}
    \centering
   \includegraphics[scale = 0.45]{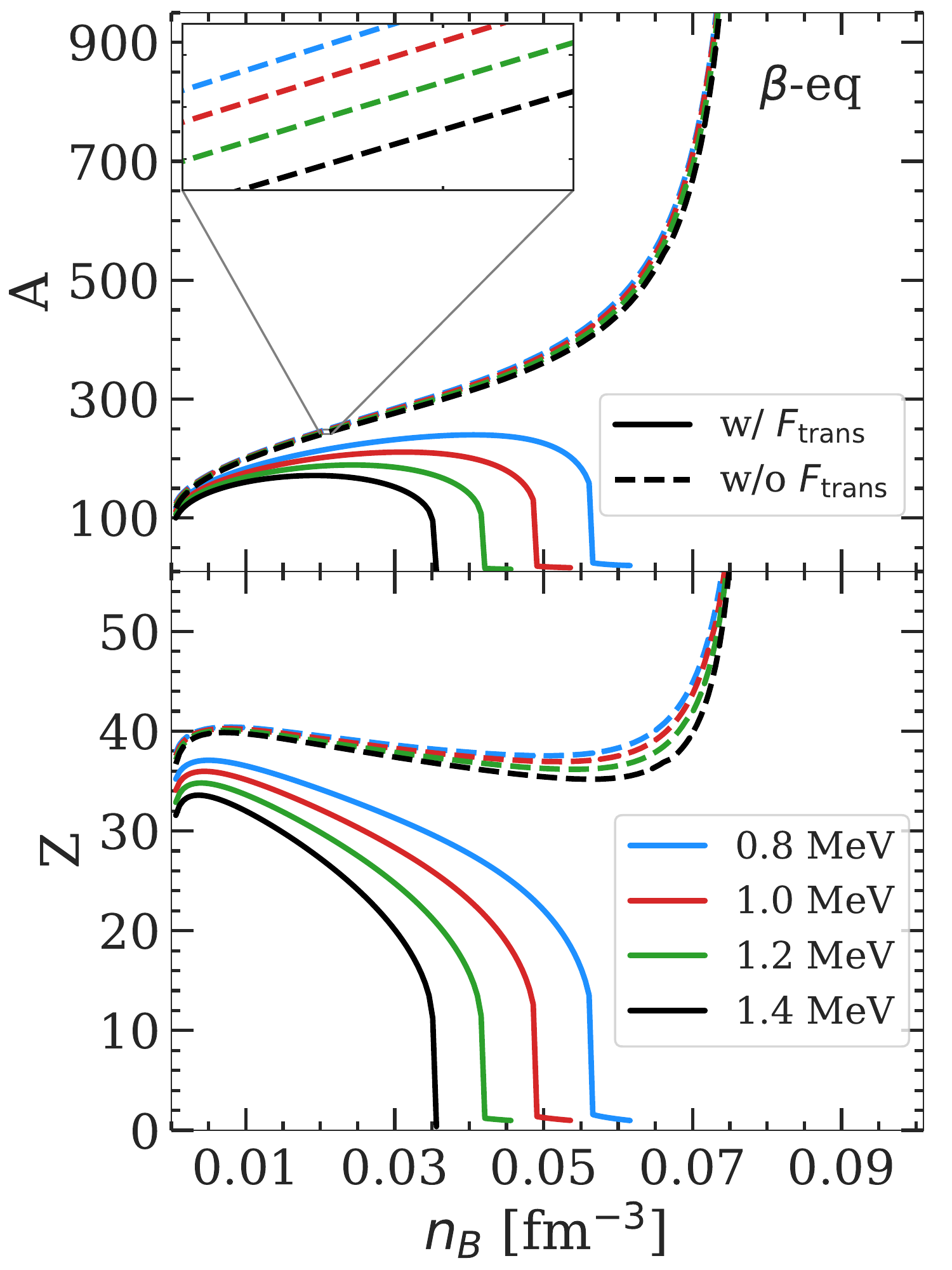}
    \caption{Cluster mass number $A$ (upper panel) and proton number $Z$ (lower panel) as a function of the baryonic density $n_B$ for the BSk24 model at four different temperatures: $k_{\rm B} T = 0.8$~MeV (blue lines), $k_{\rm B} T = 1.0 $~MeV (red lines), $k_{\rm B} T = 1.2$~MeV (green lines), and $k_{\rm B} T = 1.4$~MeV (black lines). The solid (dashed) lines are obtained when the translational energy, defined by  Eq.~(\ref{eq:ftrans}), is (not) included.}
\label{fig:betaqui_AZvsnb_bsk24}
\end{figure}
\begin{figure}
    \centering
    \includegraphics[scale = 0.45]{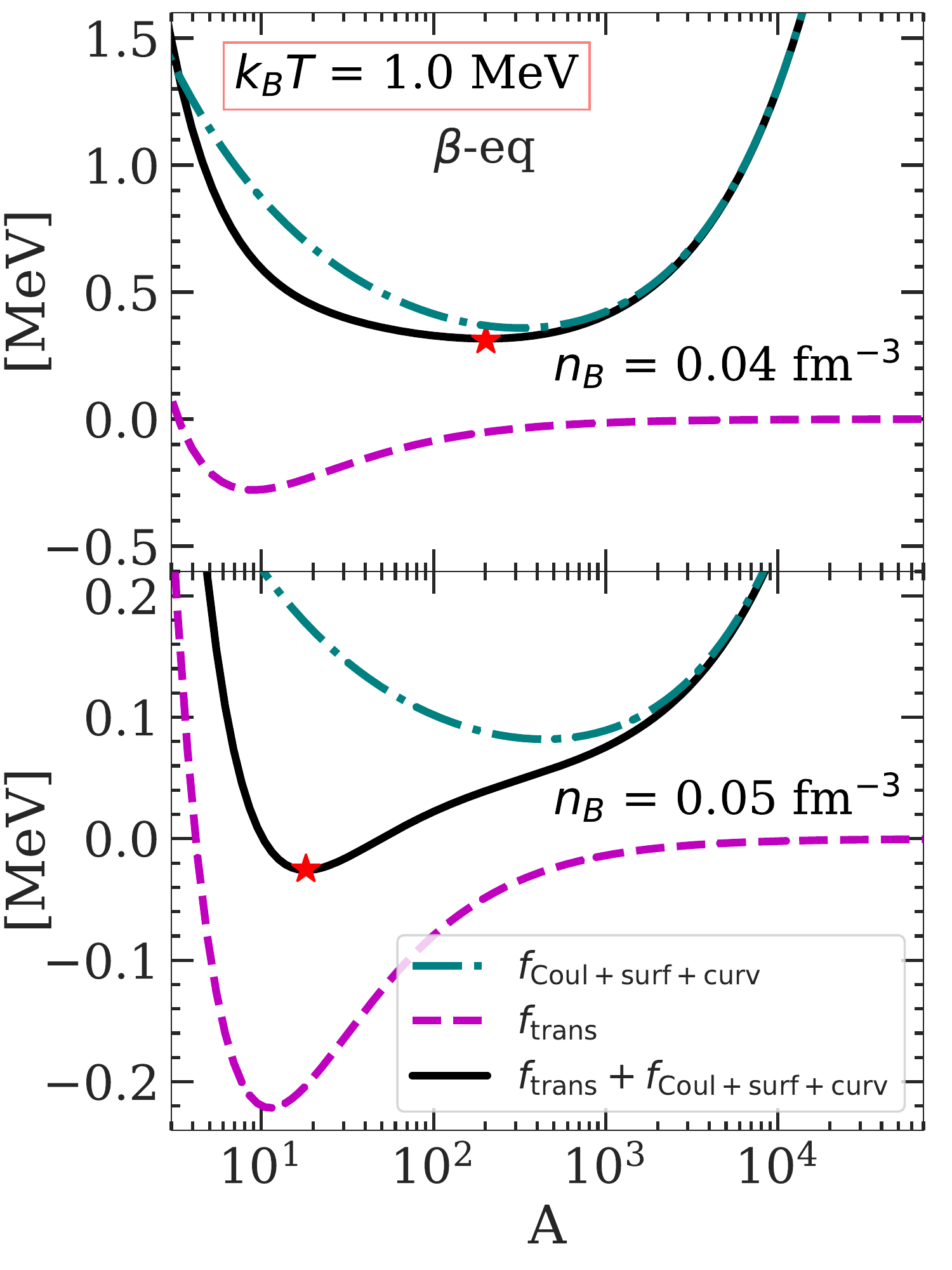}
    \caption{Coulomb, surface, and curvature energies per nucleon, $f_{\rm Coul + surf +  curv}$ (teal dash-dotted lines), translational free energy per nucleon, $f_{\rm trans}$ (violet dashed lines), and their sum (black solid lines) as a function of $A$ at $k_{\rm B} T = 1$~MeV and $n_B = 0.04$~fm$^{-3}$ (upper panel) and $n_B = 0.05$~fm$^{-3}$ (lower panel) for the BSk24 model. The red star in each panel indicates the solution for $A$ from the minimisation.}
   \label{fig:beta_energyterms_vsA}
\end{figure}

We start by considering the standard ideal-gas expression for the translational motion, Eq.~(\ref{eq:ftrans}), while the excluded-volume and effective-mass corrections are addressed later in this section.
To analyse the temperature dependence of the  properties of the inner crust, we performed the calculations employing the energy-density functional BSk24 \citep{BSK24} as an illustrative example at four selected temperature values: $k_{\rm B} T$ = 0.8, 1.0, 1.2, and 1.4 MeV.
Indeed, \citet{Carreau2020}, who calculated the melting temperature using the same BSk24 interaction model, showed that these temperatures are sufficiently high for the crust to be in the liquid phase.
Figure~\ref{fig:betaqui_AZvsnb_bsk24} shows the evolution of the cluster mass number $A$ (upper panel) and proton number $Z$ (lower panel) with the baryonic number density $n_B$ for the chosen temperatures. 
We recall that, as we  only consider spherical nuclei  here, $A$ is related to the variational variables $r_N$ and $n_i$ through the relation $A = 4 \pi r_N^3 n_i/3$, and $Z = A(1-I)/2$. 
To show the notable influence of the translational free energy $F_{\rm trans}$, Eq.~(\ref{eq:ftrans}), we show in Fig.~\ref{fig:betaqui_AZvsnb_bsk24} the results obtained including (solid lines) and neglecting (dashed lines) the $F_{\rm trans}$ term in Eq.~(\ref{eq:Fi0}).  
Let us first discuss the case without $F_{\rm trans}$ (dashed lines). 
In this case, Eq.~(\ref{eqocp:rn}) is equivalent to the equation of the Baym virial theorem \citep{bbp},  with an additional curvature term, $\mathcal{F}_{\rm surf} + 2\mathcal{F}_{\rm curv} = 2\mathcal{F}_{\rm Coul}$. From this equation, the cluster mass number $A$ can be deduced as
\begin{equation}
                A \approx  \frac{ \sigma_{\rm s}(I, T)}{ e^2 (\frac{1 - I}{2} - \frac{n_{{\rm g}p}}{n_i})^2n_i \eta_{\rm Coul}(u)} \ .
                \label{eq:A}
\end{equation}
This result is exact only if $\mathcal{F}_{\rm curv}=0$.
However, $\mathcal{F}_{\rm curv}\ll \mathcal{F}_{\rm surf}$ for all nuclei,  and Eq. (\ref{eq:A}) provides a good approximation in the whole crust. 
As $T$ increases, the surface tension decreases; see Eq.~(\ref{eq:sigmas_T}).  
Consequently, $A$ decreases with temperature, as shown in the inset of the upper panel of Fig.~\ref{fig:betaqui_AZvsnb_bsk24}. 
A similar behaviour is observed in $Z$. 
On the other hand, the denominator of the right-hand side of Eq.~(\ref{eq:A}) is a decreasing function of the baryonic density $n_B$. 
However, as nuclei become more neutron rich deeper in the crust, the surface tension $\sigma_{\rm s}(I,T)$ also decreases with $n_B$. 
As a consequence, the trend of $A$ versus $n_B$ cannot be straightforwardly deduced, and strongly depends on the detailed behaviour of $\sigma_{\rm s}$, that is, on the nuclear model. 
From Fig.~\ref{fig:betaqui_AZvsnb_bsk24}, one can see that for the considered temperatures and functional, the behaviour of the denominator prevails, hence the increase in cluster mass.  
For $Z$, this is not always the case because the isospin $I$ also increases with $n_B$.
Taking into account the translational motion $F_{\rm trans}$ (solid lines in Fig. \ref{fig:betaqui_AZvsnb_bsk24}) significantly reduces both the mass and proton numbers of the cluster.
Specifically, at high densities, the cluster completely dissolves, that is, $Z < 1$. 
Analogously to the calculation without the translational term, the cluster size decreases with temperature. 
However, the density at which the transition from inhomogeneous to homogeneous
matter occurs is much lower than that obtained without $F_{\rm trans}$.
In particular, if we neglect (include) $F_{\rm trans}$, for all the considered temperatures, the crust--core transition occurs at around $n_{\rm CC} \approx 0.08$~fm$^{-3}$ ($n_{\rm CC} \approx 0.04 - 0.06$~fm$^{-3}$), as can be seen from Table~\ref{table1:cc_transition}. 

\begin{table}
\centering
\caption{Crust--core transition density, $n_{\rm CC}$, and pressure, $P_{\rm CC}$, for the BSk24 functional at different temperatures with (w/) and without (w/o) the inclusion of the translational free energy $F_{\rm trans}$.}
\label{table1:cc_transition}
\setlength{\tabcolsep}{6pt}
\begin{tabular}{|cl|cccc|}
\hline
\hline
\multicolumn{2}{|c|}{$k_{\rm B}T$}             
& 0.8 & 1.0 & 1.2 & 1.4 \\
\multicolumn{2}{|c|}{(MeV)}&  &  &  &  \\
\hline
\multicolumn{1}{|c|}{$n_{\rm CC}$} & w/ $F_{\rm trans}$  &  0.062   &  0.054   & 0.046    &  0.036   \\ \cline{2-6} 
\multicolumn{1}{|c|}{(fm$^{-3}$)}  & w/o $F_{\rm trans}$ &  0.079     &  0.079   &  0.078   &  0.078    \\ 
\hline
\multicolumn{1}{|c|}{$P_{\rm CC}$} & w/ $F_{\rm trans}$  & 0.160    &  0.133   &   0.111   &  0.078  \\ \cline{2-6} 
\multicolumn{1}{|c|}{(MeV/fm$^3$)} & w/o $F_{\rm trans}$ &  0.260    &    0.257   &   0.255   &  0.256   \\
\hline
\hline
\end{tabular}%
\end{table}

To understand this behaviour, we observe that the size of the cluster is determined by the minimisation of the free energy per nucleon with respect to $r_N$ (or equivalently $A$), Eq.~(\ref{eqocp:rn}). 
From the definition of $F_i$ in Eq.~(\ref{eq:Fi0}), Eq.~(\ref{eqocp:rn}) can be written as
\begin{equation}
        \frac{\partial (f_{\rm Coul + surf +  curv} + f_{\rm trans})}{\partial r_N} = 0.
        \label{eq:rn1}
\end{equation}  
This equation shows that we can have heavy or light clusters depending on the competition between $f_{\rm Coul + surf +  curv}$, which favours the former (as we show above), and $f_{\rm trans}$, which favours the latter. 
Indeed, it is easy to show that $f_{\rm trans}$ is minimised at
        \begin{equation}
                A \approx A_0 =  \exp \left[ \frac{2}{5}\ln\left (C_0\frac{n_i u}{T^{3/2}}\right ) + \frac{3}{5}\right] \ ,
                \label{eq:Amin}
        \end{equation}
 where  $C_0 = \hbar^3/g_s \left( 2\pi/m_n k_{\rm B}  \right) ^{3/2}$.

\begin{figure}
    \centering
    \includegraphics[scale = 0.45]{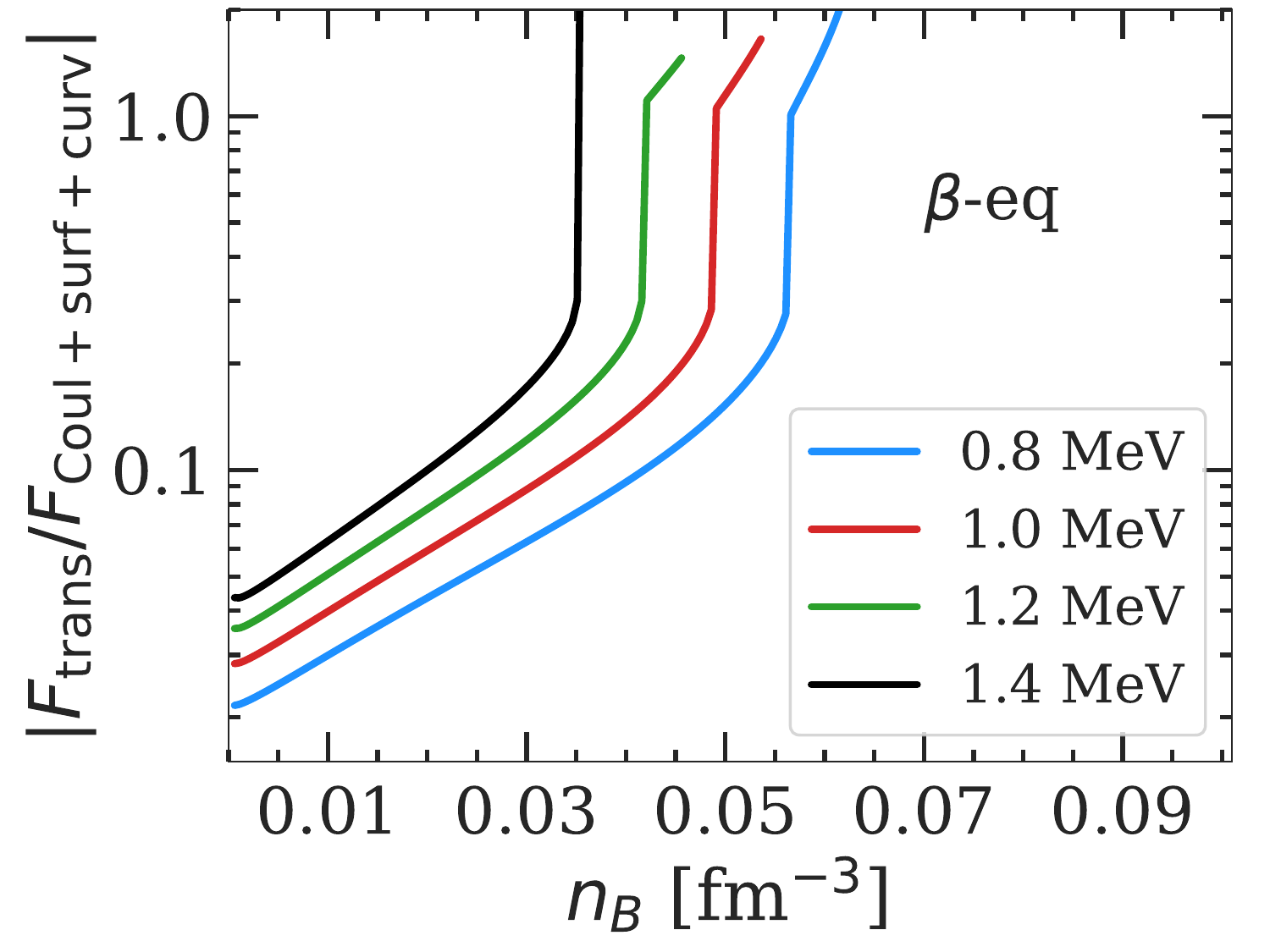}
    \caption{Absolute value of the ratio of the translational free energy to the sum of surface, curvature, and Coulomb free energies as a function of the baryonic density $n_B$ for the BSk24 model at four different temperatures: $k_{\rm B} T = 0.8$~MeV (blue line), $k_{\rm B} T = 1.0 $~MeV (red line), $k_{\rm B} T = 1.2$~MeV (green line), and $k_{\rm B} T = 1.4$~MeV (black line).} 
\label{fig:betaqui_fenergy_ratio_bsk24}
\end{figure}

Replacing realistic values for $n_i$ and $u$ in Eq.~(\ref{eq:Amin}), we find that  $A_0 <30$ for temperatures equal to or above 1~MeV. 
This point is further illustrated in Fig.~\ref{fig:beta_energyterms_vsA}, in which we plot $f_{\rm Coul + surf +  curv}$ (teal dash-dotted lines), $f_{\rm trans}$ (violet dashed lines), and their sum (black solid lines) as a function of $A$ at $k_{\rm B} T = 1$~MeV for two selected densities: $n_B = 0.04$~fm$^{-3}$ (upper panel) and $n_B = 0.05$~fm$^{-3}$ (lower panel). 
The other variables ($I$, $n_i$, $n_p$, $n_{{\rm g}n}$, $n_{{\rm g}p}$, $n_e$) are fixed to the values obtained from the minimisation at the given ($n_B, T$). 
From this figure, we can see that, indeed, $f_{\rm Coul + surf +  curv}$ is minimised for large clusters ($A > 100$), whereas $f_{\rm trans}$ is minimised for very small ones, $A \approx 10$. 
At higher density, as matter becomes more neutron rich, both the Coulomb and the surface energies per baryon decrease. 
Thus, the impact of the translational motion prevails. 
Moreover, $f_{\rm trans}$ clearly dominates at high temperatures.
Consequently, the net effect is that the translational free energy is dominant at high temperatures and densities, as shown in Fig. \ref{fig:betaqui_fenergy_ratio_bsk24}. 
This explains why small clusters are found in this regime in Fig.~\ref{fig:betaqui_AZvsnb_bsk24} when the translational energy is accounted for in the minimisation procedure. 
\begin{figure}
    \centering
    \includegraphics[scale = 0.38]{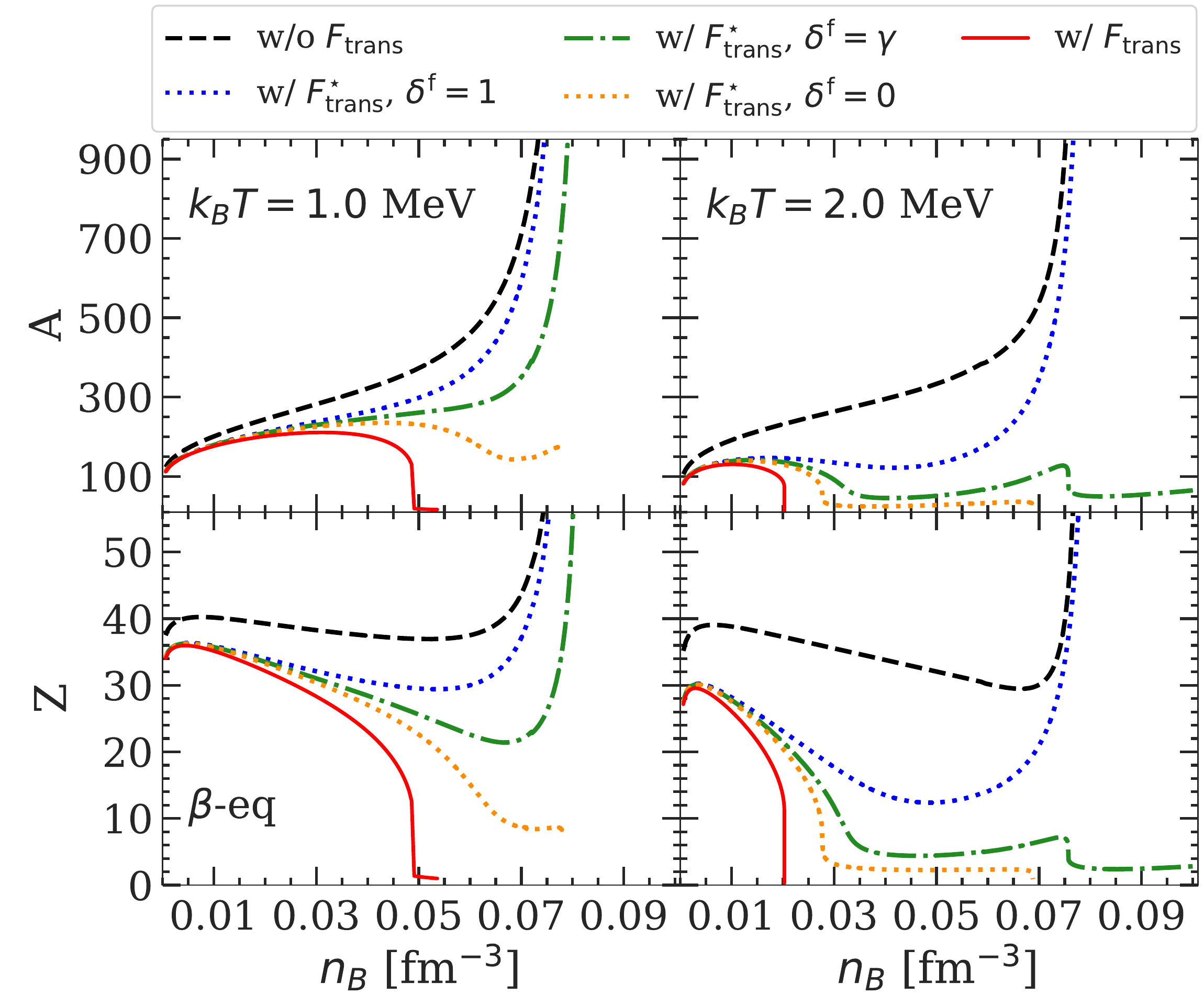}
    \caption{Cluster mass number $A$ (upper panels) and proton number $Z$ (lower panels) as a function of the baryonic density $n_B$ for the BSk24 model at two different temperatures: $k_{\rm B} T = 1.0$~MeV (left panels) and $k_{\rm B} T = 2.0$~MeV (right panels). Different prescriptions for the translational free energy are considered. See text for details.}
    \label{fig:bsk24_betaequi_modifiedftrans_AZ_3meff-linearcalex}
\end{figure}
\begin{figure}
    \centering
    \includegraphics[scale = 0.38]{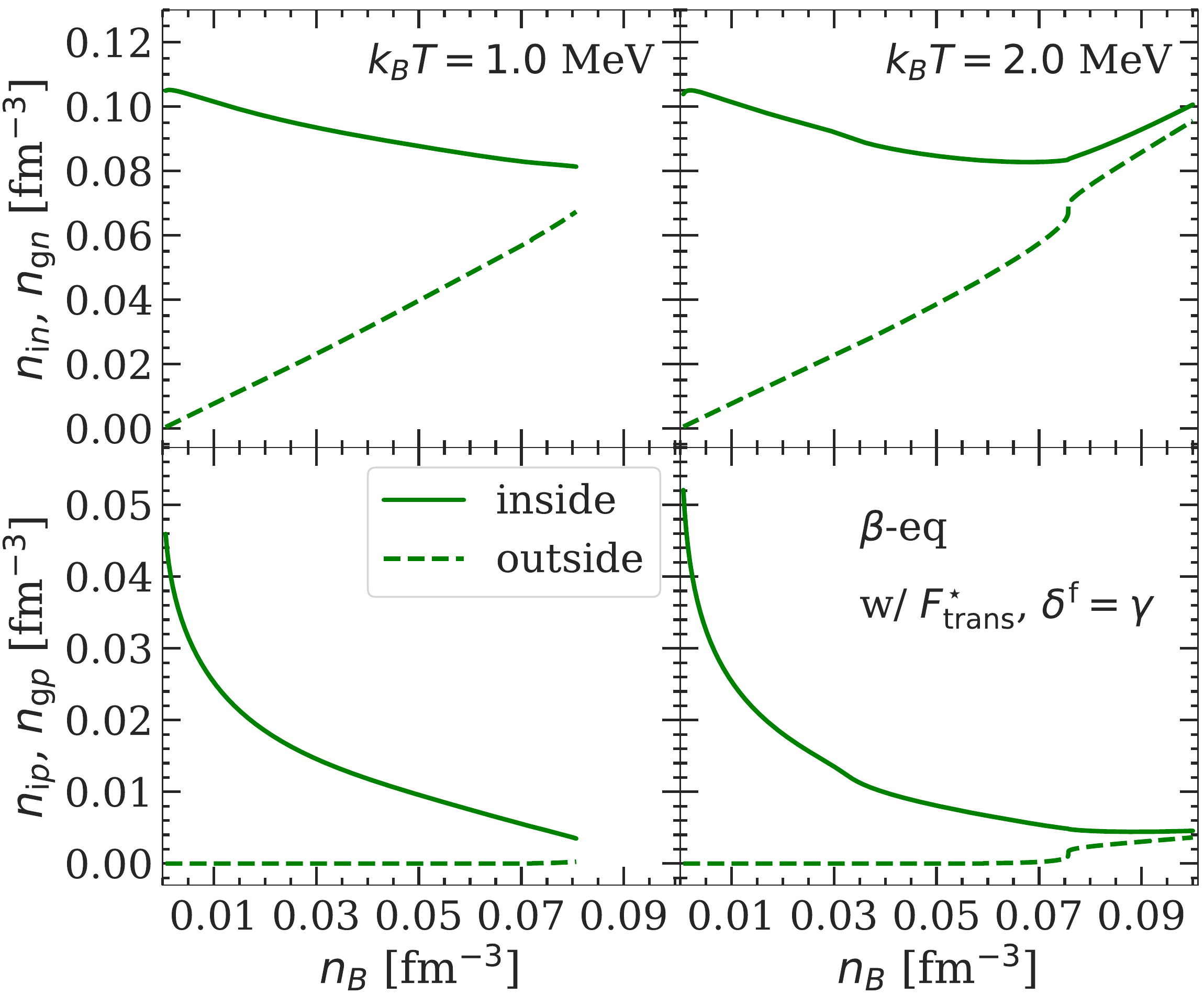}
    \caption{Neutron (upper panels) and proton (lower panels) densities inside (solid lines) and outside (dashed lines) the cluster as a function of the baryonic density $n_B$ for the BSk24 model at two different temperatures: $k_{\rm B} T = 1.0$~MeV (left panels) and $k_{\rm B} T = 2.0$~MeV (right panels). The effective translational free energy  $F^{\star}_{\rm trans}$ is included, with $\delta^{\rm f} = \gamma$. }
    \label{fig:bsk24_betaequi_modifiedftrans_densities_deltaf-eqgamma-linearscalex}
\end{figure}
\begin{figure*}
    \centering
   \includegraphics[scale = 0.45]{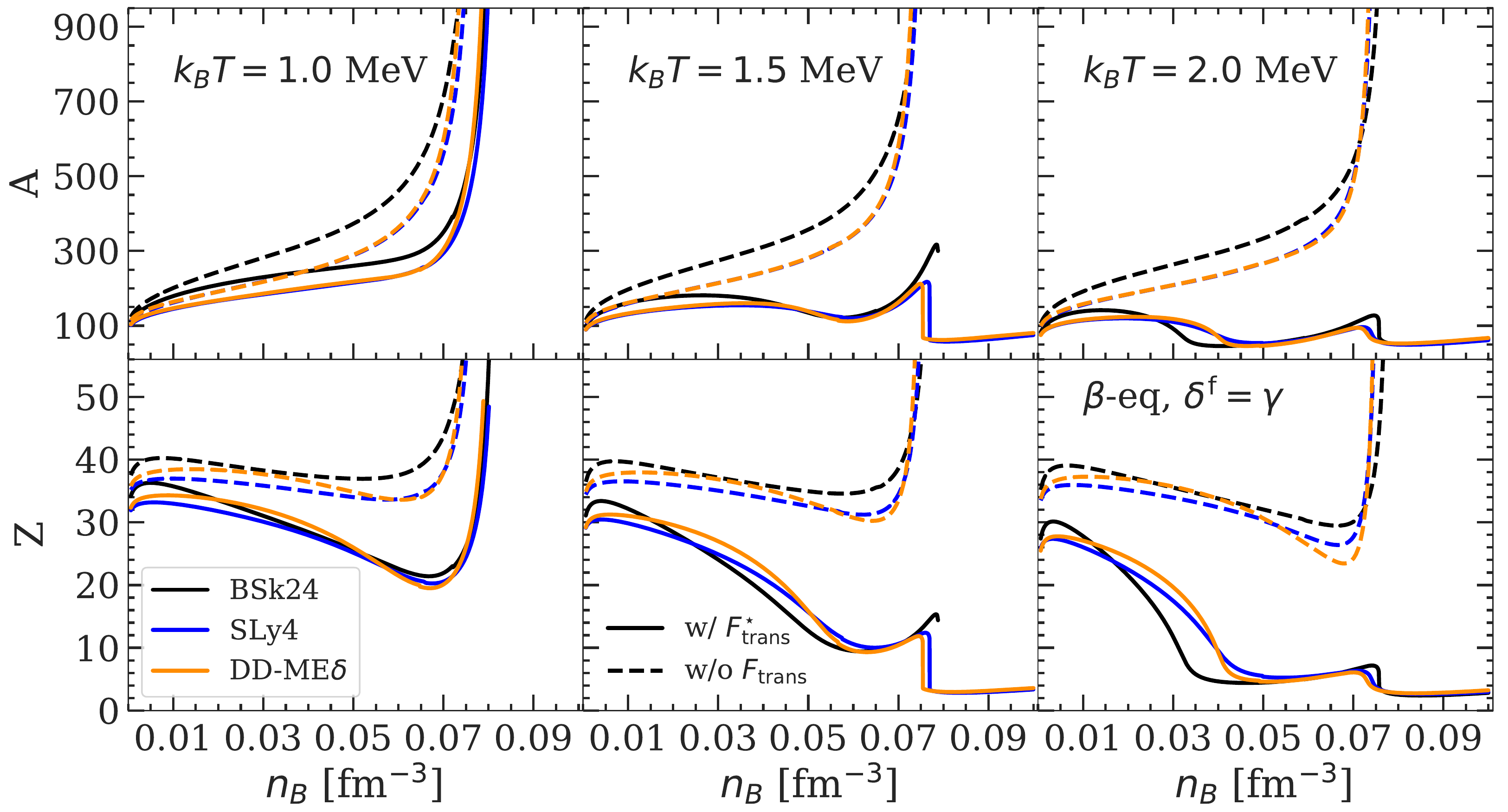}
    \caption{Cluster mass number $A$ (upper panels) and proton number $Z$ (lower panels) as a function of the baryonic density $n_B$ for the BSk24 (black lines), SLy4 (blue lines), and DD-ME$\delta$ (orange lines) models at three different temperatures: $k_{\rm B} T = 1.0$~MeV (left panels), $k_{\rm B} T = 1.5$~MeV (middle panels), and $k_{\rm B} T = 2.0$~MeV (right panels). The solid (dashed) lines are obtained with (without) the effective translational energy ${F}_{\rm trans}^\star$, Eq.~(\ref{eq:ftrans_eff}), with $\delta^{\rm f} = \gamma$. }
    \label{fig:betaequi_modifiedftrans_AZ}
\end{figure*}
\begin{figure}
    \centering
     \includegraphics[scale = 0.45]{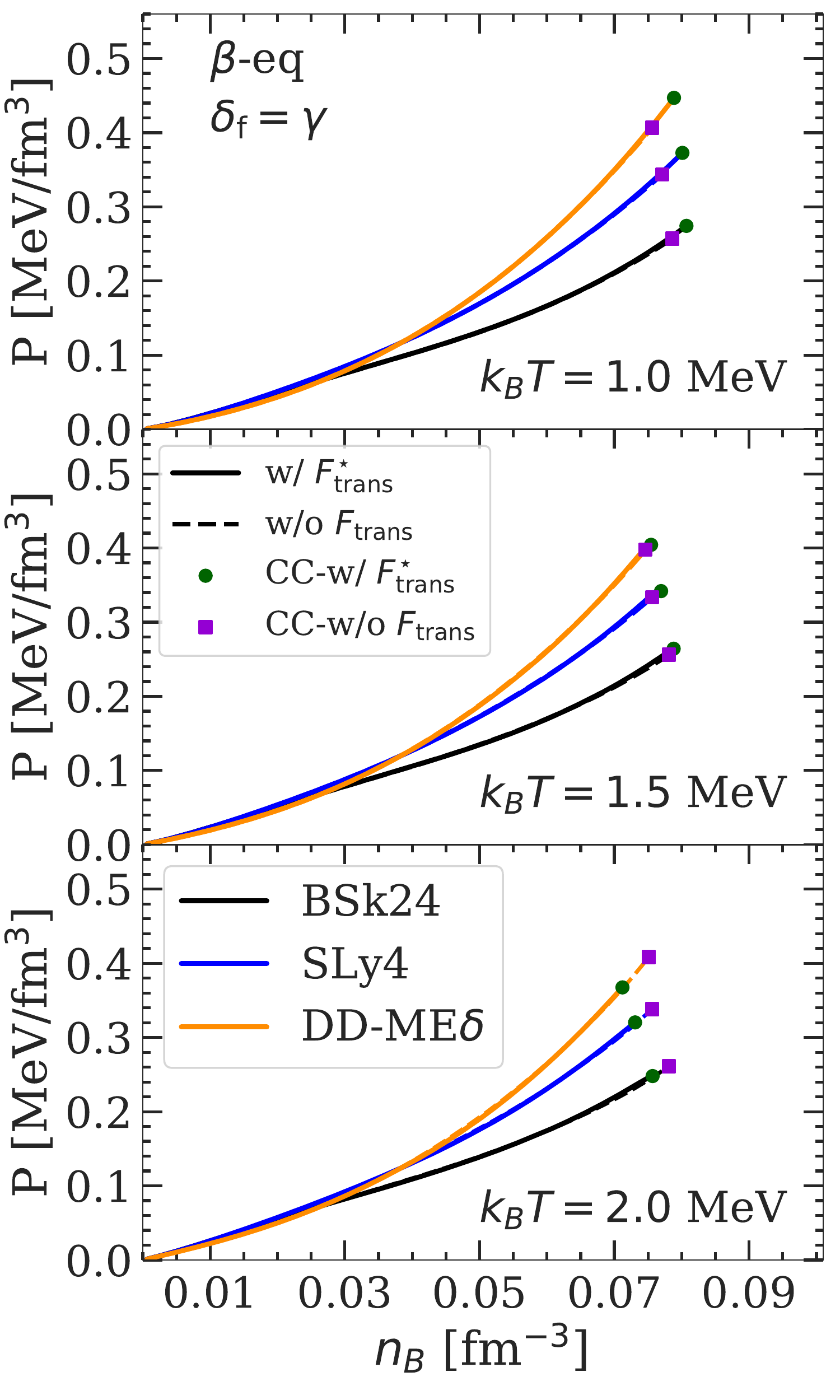}
    \caption{Same as Fig.~\ref{fig:betaequi_modifiedftrans_AZ}, but for pressure $P$ vs.  baryonic number density $n_B$. The green dots (violet squares) mark the crust--core transition with (without) ${F}_{\rm trans}^\star$. }
    \label{fig:betaequi_modifiedftrans_EoS}
\end{figure}

The surprising results of Figs.~\ref{fig:betaqui_AZvsnb_bsk24}-\ref{fig:betaqui_fenergy_ratio_bsk24} are obtained using the oversimplified ideal-gas expression in Eq.~(\ref{eq:ftrans}), which is only justified at rather high temperature and relatively low density; see Fig.~\ref{fig:ratio-lambda-Vws}.
One would expect the inclusion of the cluster effective mass and the reduction of the available volume for the translational motion to reduce the impact of the centre-of-mass motion. 
This is shown in Fig.~\ref{fig:bsk24_betaequi_modifiedftrans_AZ_3meff-linearcalex}, where the cluster mass number $A$ and proton number $Z$ are plotted as a function of baryonic number density $n_B$ for the BSk24 model at $k_{\rm B}T = 1.0$~MeV (left panels) and $k_{\rm B}T = 2.0$~MeV (right panels). For all the considered values of $\delta^{\rm f}$, including  ${F}_{\rm trans}^\star$(Eq.~(\ref{eq:ftrans_eff})) still noticeably shrinks the size of the cluster with respect to the case where the collective degrees of freedom are neglected. However, the effect is not as dramatic as when the ideal-gas expression in Eq.~(\ref{eq:ftrans}) is used (see also Fig.~\ref{fig:betaqui_AZvsnb_bsk24}).
The latter is true both when $M_i^\star < M_i$, that is, when $\delta^{\rm f} > 0$, and when $M_i^\star > M_i$, that is, in the case where $\delta^{\rm f} = 0$. 
As one may expect, the influence of the translational free energy becomes larger with increasing temperature, as can be seen by comparing the left and right panels. 
The effect is also larger with decreasing $\delta^{\rm f}$. 
As already mentioned, among the three prescriptions of the ion effective mass discussed in Sect.~\ref{sec:hydro}, the most realistic one would be when all neutrons in the continuum states participate in the flow, that is, $\delta^{\rm f} = \gamma$.
Therefore, in the following, we  only show our results for this case, that is, we employ Eq.~(\ref{eq:m*-deltamin}) in $F_{\rm trans}^\star$ (Eq.~(\ref{eq:ftrans_eff})).

Interestingly, in the case where $\delta^{\rm f} = \gamma$ (green dash-dotted line in Fig.~\ref{fig:bsk24_betaequi_modifiedftrans_AZ_3meff-linearcalex}), we observe a transition to small proton-mass clusters, $Z \sim 2$, at $k_BT = 2.0$~MeV (right panels) and $n_B > 0.075$~fm$^{-3}$, and this configuration persists even at $n_B> 0.1$~fm$^{-3}$. 
As shown in Fig.~\ref{fig:bsk24_betaequi_modifiedftrans_densities_deltaf-eqgamma-linearscalex}, the neutron and proton cluster densities (solid lines), $n_{{\rm i}n} = n_i(1 + I)/2$ and $n_{{\rm i}p} = n_i(1 - I)/2$, of these light configurations are very similar to those of the surrounding homogeneous nuclear medium (dashed lines).
From a physical point of view, these small-amplitude and small-wavelength inhomogeneities might correspond to correlations or resonant states that are expected to exist in nuclear matter but are not captured in the homogeneous matter mean-field approximation. 
However, in this density region, there is no clear distinction between inhomogeneous and homogeneous matter. 
Therefore, we identify this transition to low-$Z$ clusters as the interface between the crust and the core.

In Fig.~\ref{fig:betaequi_modifiedftrans_AZ}, we plot the cluster mass number $A$ and proton number $Z$ as a function of baryonic number density $n_B$ for three temperature values: $k_{\rm B} T = 1.0$~MeV (left panels), $1.5$~MeV (middle panels), and $2.0$~MeV (right panels) for the BSk24 \citep{BSK24} (black lines), SLy4 \citep{SLy4} (blue lines), and DD-ME$\delta$ \citep{DDMEd} (orange lines) models. 
The $A$ and $Z$ obtained with (without) the translational energy $F^\star_{\rm trans}$ are shown with solid (dashed) lines.
While the results are almost model-independent, they vary significantly with temperature. 
For all the considered models, with the most realistic expression that we consider in this work, ${F}_{\rm trans}^\star$, at 1~MeV, very heavy clusters ($A> 400$) are favoured until the transition to the core, whereas at higher temperatures, the crust--core transition occurs at lower densities.
The latter behaviour can be expected due to the increase of the translational degrees of freedom. 

To better understand this finding, we notice that the mass number $A^{\star}_0$ at which ${F}_{\rm trans}^\star/A$ is minimised has the exact same functional form as in Eq.~(\ref{eq:Amin}), that is
\begin{equation}
                  {A}_0^\star =  \exp \left[ \frac{2}{5}\ln\left (C_0\frac{n_i u}{T^{3/2}g(u,\gamma)}\right )  
+ \frac{3}{5}\right],
                \label{eq:Amin_tilde}
\end{equation}
where 
\begin{equation}
     g(u,\gamma) = \left (\sqrt{1-\gamma}(1-u^{1/3})\right)^{3}.
                \label{eq:C0_tilde}
\end{equation}
The behaviour of $A_0^\star$ therefore depends on the competition between the $T^{3/2}$ and the $g(u,\gamma)$ factors.
As $n_B$ increases, the ratio $\gamma$ between the density outside and inside the ion increases and approaches 1 near the crust--core transition, thus reducing the effect of the translational term. 
If the volume fraction occupied by the ion $u$ also increases (approaching 1), which is the case at relatively low temperatures, $1/g$ can be a very large number, and large clusters are still favoured even if the translational free energy is accounted for (see left panels of Fig.~\ref{fig:betaequi_modifiedftrans_AZ}). 
However, at higher temperatures, that is $k_BT \gtrsim 1.5$~MeV, we find that the translational effect dominates, and the cluster volume fraction decreases near the crust--core transition, which leads to the appearance of small clusters when ${F}_{\rm trans}^\star$ is included.

The impact of the excluded-volume approach on the transition from clusterised to homogeneous matter was also discussed by \citet{Typel2016}.
This author noticed a strong increase (and even divergence) of the pressure, which resulted in the transition to uniform matter when the excluded-volume mechanism was implemented, and proposed a solution, namely a generalisation of the excluded-volume approach.
In our case, this same divergence that is brought by the term $(1-u^{1/3})^{-3}$ in Eq.~(\ref{eq:Amin_tilde}) is moderated by finite-temperature effects, yielding instead the appearance of lighter, essentially unbound clusters.

The effect of the modified translational free energy, ${F}_{\rm trans}^\star$, on the compositions ($A$, $Z$) is significant at all densities in the proto-NS inner crust. 
Specifically, at the highest temperature considered, $k_{\rm B} T = 2.0$~MeV, the reduction in $A$ and $Z$ with respect to a calculation where the translational energy is neglected or excluded from the variational equation with respect to the cluster size, as in the works by \citet{lattimer1991} and \citet{Shen2011}, is of the order of $>400$ in $A$ and $>20$ in $Z$.
Therefore, for calculations of the (liquid) inner crust where an accurate description of the composition is needed, we believe that the translational free energy should be taken into account. 
In contrast to the ideal-gas approximation, the inclusion of in-medium effects in the translational free energy does not lead to an early dissolution of the clusters, as can be seen by comparing Figs.~\ref{fig:betaqui_AZvsnb_bsk24} and \ref{fig:betaequi_modifiedftrans_AZ}. 
Instead, at all the considered temperatures, clusters are favoured over homogeneous nuclear matter up to $n_B \gtrsim 0.07$~fm$^{-3}$.

On the other hand, for most cases, the impact of the translational free-energy term is reduced on more global quantities like the equation of state. 
This is demonstrated in Fig.~\ref{fig:betaequi_modifiedftrans_EoS}, where the total pressure $P$ is plotted as a function of the baryon number density $n_B$ in the inner crust. 
We can see that, for all the considered models, including the in-medium modified translational free energy ${F}_{\rm trans}^\star$ yields almost identical results to the case where it is not included; see the solid and dashed curves in Fig.~\ref{fig:betaequi_modifiedftrans_EoS}, which are almost indistinguishable.
The crust--core interface is indicated by green dots (violet squares) for the case where ${F}_{\rm trans}^\star$ is (not) taken into account. 
The crust--core transition was determined 
by comparing the WS free-energy density of the inhomogeneous crust to that of homogeneous (or quasi-homogeneous)  matter at beta equilibrium (see also Fig.~\ref{fig:betaequi_modifiedftrans_AZ}). 
At $k_BT = 1.0$~MeV (top panel), the translational degrees of freedom favour clusters over homogeneous matter by lowering the free-energy density of the system. 
Thus, the crust--core transition is moved to higher density (and pressure) when this term is accounted for. 
On the other hand, the situation is reversed for $k_BT = 1.5$~MeV (middle panel) and $2.0$~MeV (bottom panel).
This is understood from the fact that the most favourable clusters decrease in size as the temperature increases, leading to a progressive `melting' in the surrounding medium.
Therefore, the resulting crust--core transition is very close to or even lower than that obtained when neglecting the translational-free-energy contribution. 
Values of the crust--core transition density and pressure are reported in Table \ref{table2:cc_transition_3models}.

\begin{table*}
\centering
\caption{Crust--core transition density, $n_{\rm CC}$, and pressure, $P_{\rm CC}$, for the BSk24, SLy4, and DD-ME$\delta$ models at different temperatures with (w/) and without (w/o) the inclusion of the translational free energy ${F}_{\rm trans}^\star$ for the case $\delta^{\rm f} = \gamma$.}
\label{table2:cc_transition_3models}
\setlength{\tabcolsep}{6pt}
\begin{tabular}{|cl|ccc|ccc|ccc|}
\hline
\hline
\multicolumn{2}{c}{Model} & \multicolumn{3}{c}{BSk24} &\multicolumn{3}{c}{SLy4} & \multicolumn{3}{c}{DD-ME$\delta$} \\
\hline
\multicolumn{2}{|c|}{$k_BT$} & 1.0 & 1.5 & 2.0 &  1.0 & 1.5 & 2.0 & 1.0 & 1.5 & 2.0 \\
\multicolumn{2}{|c|}{(MeV)}&  &  &  & & & & & &  \\
\hline
\multicolumn{1}{|c|}{$n_{\rm CC}$} & w/ $F^{\star}_{\rm trans}$ &   0.081      &    0.079  &  0.076
&  0.080 & 0.077 & 0.073
&  0.079 & 0.075 & 0.071 \\ \cline{2-11} 
\multicolumn{1}{|c|}{(fm$^{-3}$)}  & w/o $F^{\star}_{\rm trans}$ &   0.079  & 0.078    &  0.078   &  0.077   & 0.076 & 0.076 
& 0.076 & 0.075& 0.075 \\ 
\hline
\multicolumn{1}{|c|}{$P_{\rm CC}$} & w/ $F^{\star}_{\rm trans}$  &  0.274   &  0.264  & 0.248  
& 0.373  &  0.342 & 0.320
& 0.447 &  0.404 &  0.368 \\ \cline{2-11} 
\multicolumn{1}{|c|}{(MeV/fm$^3$)} & w/o $F^{\star}_{\rm trans}$ &  0.257    & 0.256     &  0.261    &  0.344  & 0.334 & 0.338 
& 0.407 & 0.398&  0.408\\
\hline
\hline
\end{tabular}%
\end{table*}

\section{Results at fixed total proton fraction}
\label{sec:result_fix_yp}
\begin{figure}
    \centering
   \includegraphics[scale = 0.45]{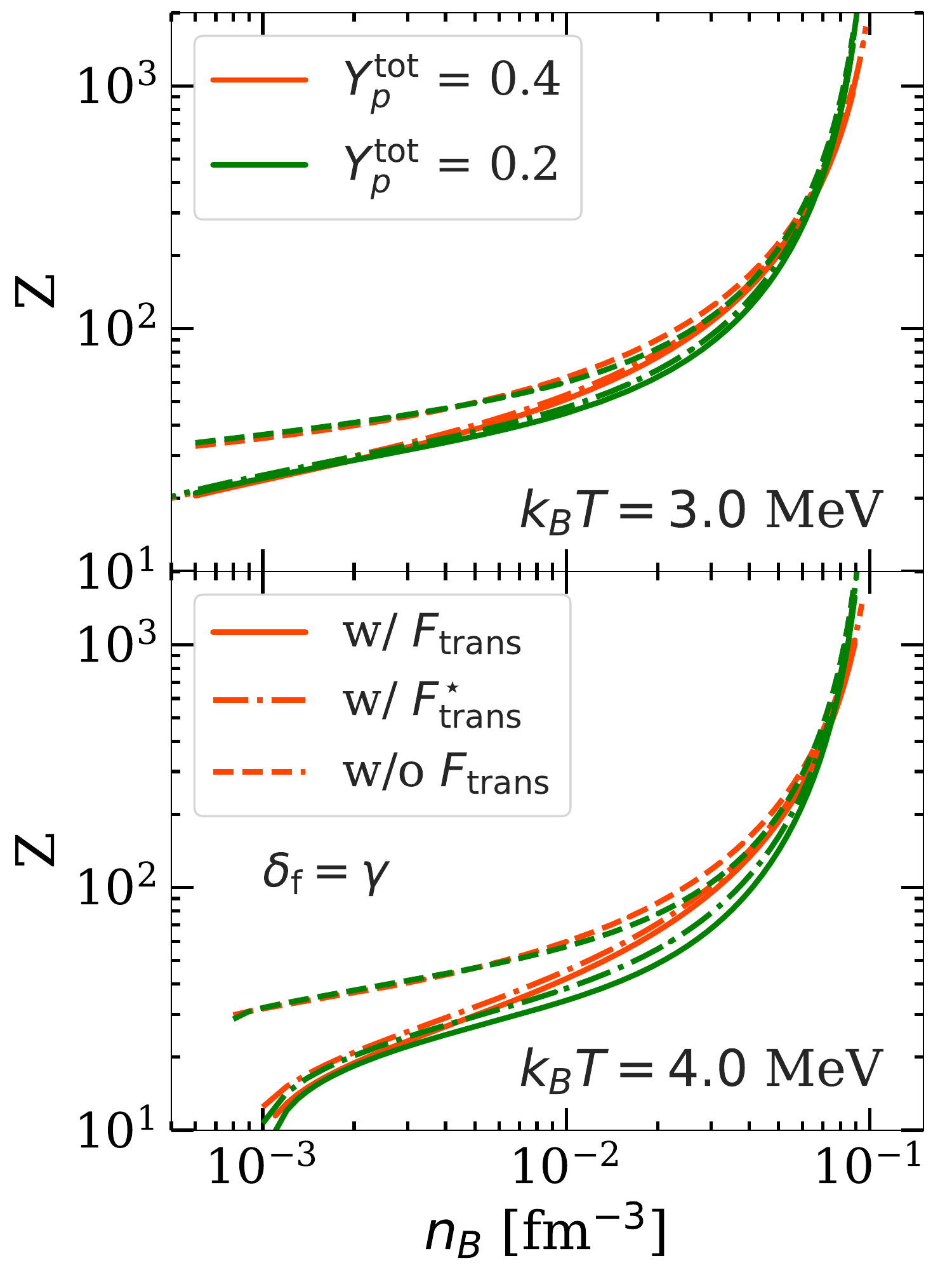}
    \caption{Proton number $Z$ as a function of the baryonic density $n_B$ for the BSk24 model at $Y_p^{\rm tot}$ = 0.4 (red lines) and $Y_p^{\rm tot}$ = 0.2 (green lines) for $k_{\rm B} T = 3.0$~MeV (upper panel) and $k_{\rm B}T = 4.0$~MeV (lower panel). Different prescriptions for the translational free energy are considered. See text for details.}
    \label{fig:fixyp_Zvsnb_bsk24}
\end{figure}

\begin{figure}
    \centering
\includegraphics[scale = 0.45]{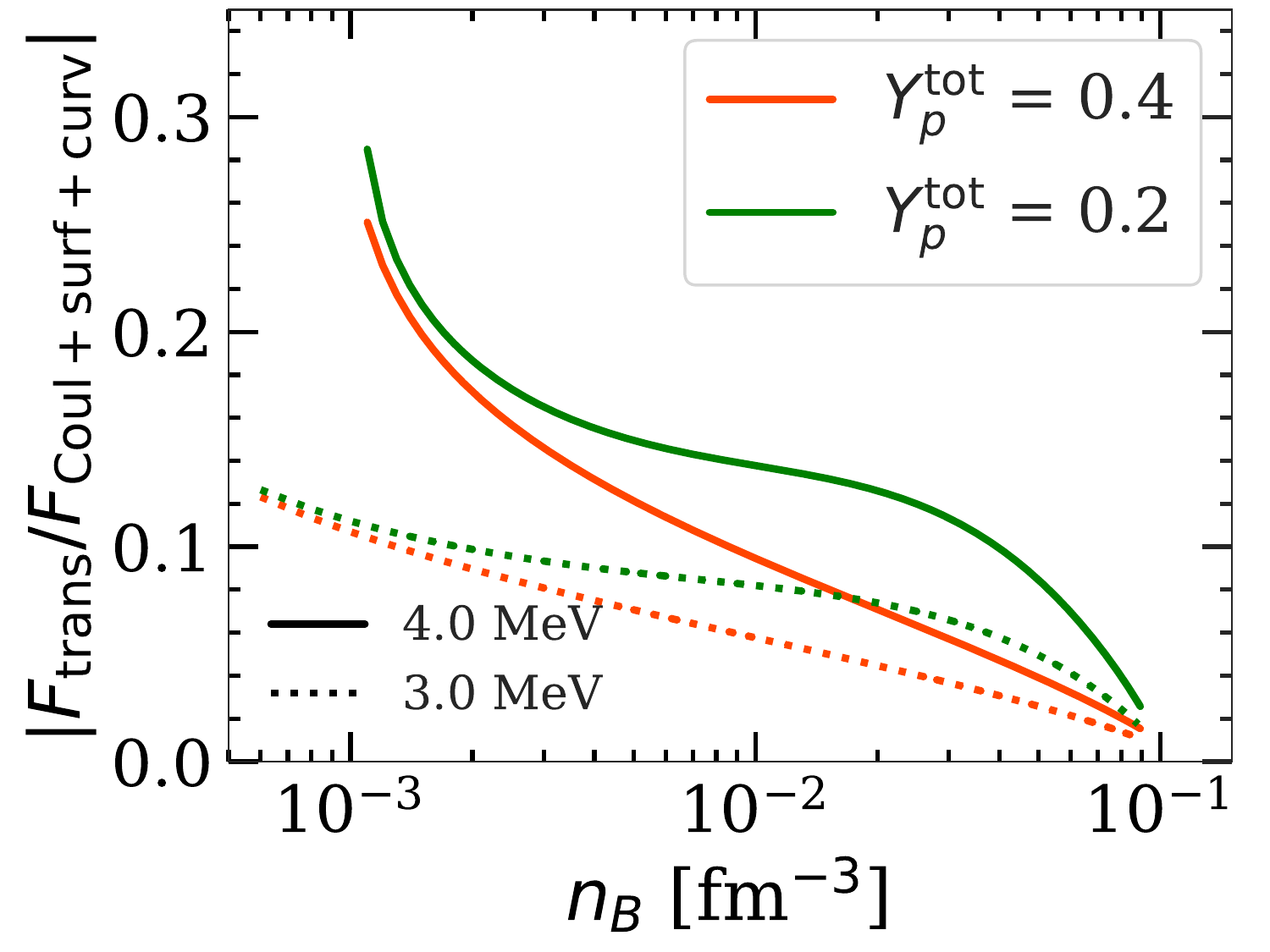}
    \caption{Absolute value of the ratio of the translational energy to the sum of surface, curvature, and Coulombs energies as a function of the baryonic density $n_B$ for the BSk24 model at $k_{\rm B} T = 4.0$~MeV (solid lines) and $k_{\rm B}T = 3.0$~MeV (dotted lines), for $Y_p^{\rm tot} = 0.4$ (red lines) and $Y_p^{\rm tot}= 0.2$ (green lines).}
\label{fig:fixyp_fenergy_ratio_bsk24}
\end{figure}
\begin{figure}
    \centering
    \includegraphics[scale = 0.43]{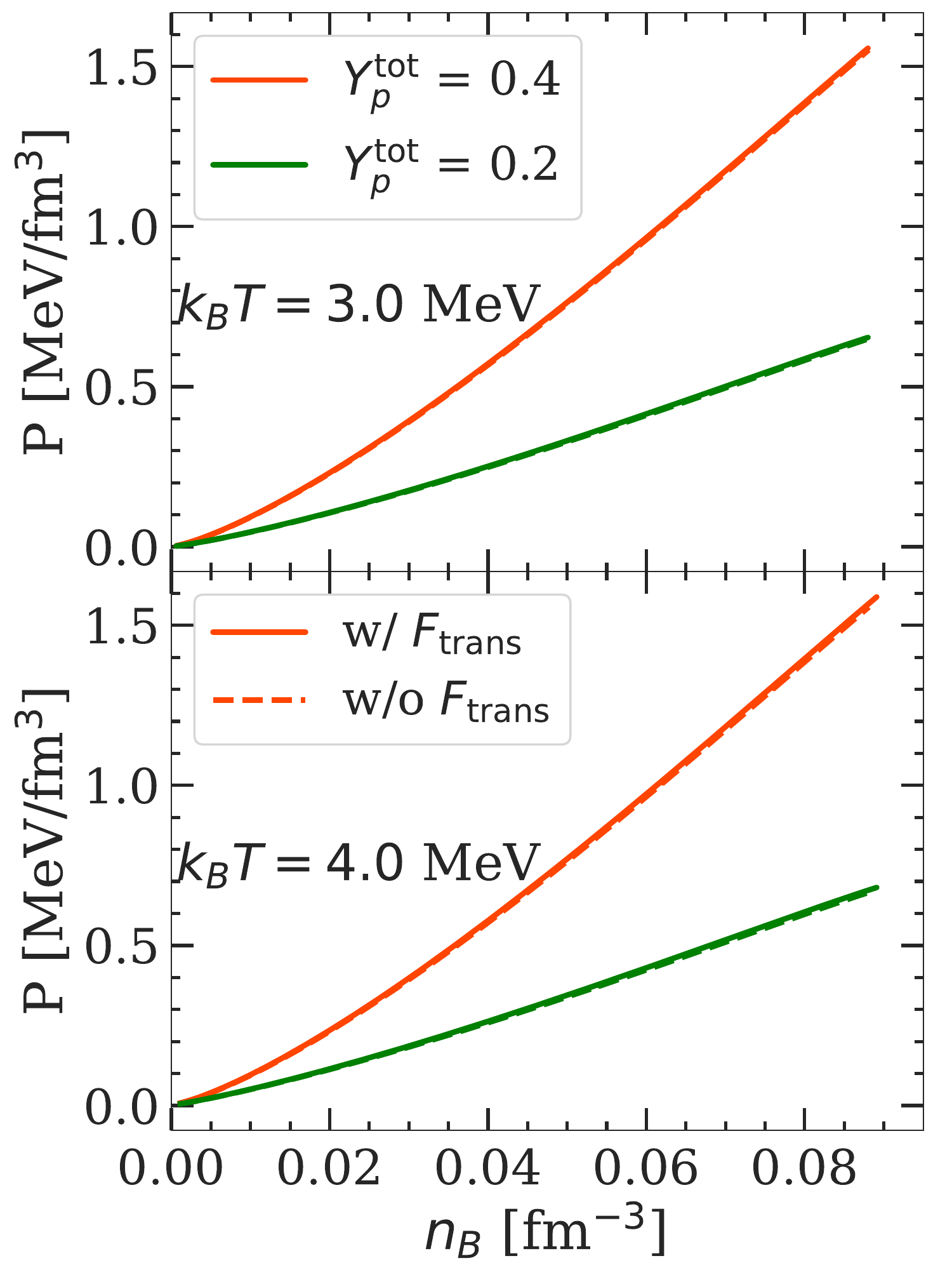}
    \caption{Same as in Fig.~\ref{fig:fixyp_Zvsnb_bsk24} but for pressure $P$ vs. baryonic number density $n_B$.}   \label{fig:fixyp_EoS_bsk24}
\end{figure}

At a fixed total proton fraction $Y_p^{\rm tot} = n_e/n_B$, Eq.~(\ref{eqocp:betaequi}) associated to beta equilibrium no longer holds, and the equilibrium state of matter for each given thermodynamic condition ($n_B$, $T$, $Y_p^{\rm tot}$) is determined by solving the system of four variational equations, Eqs.~(\ref{eqocp:rn})-(\ref{eqocp:Iandnp}), together with the equations of baryon number conservation, Eq.~(\ref{eq:baryon_conservation}), and charge neutrality, Eq.~(\ref{eq:charge}). 

Figure~\ref{fig:fixyp_Zvsnb_bsk24} displays the cluster proton number $Z$ as a function of $n_B$ at $k_{\rm B}T = 3.0$~MeV (upper panel) and $k_{\rm B} T = 4.0$~MeV (lower panel) at two selected total proton fractions: $Y_p^{\rm tot} = 0.4$ (red lines) and $Y_p^{\rm tot} = 0.2$ (green lines). 
Results are shown for the case where the translational free energy is neglected (dashed lines) and for the two different prescriptions for the translational-free-energy term considered in this work: $F_{\rm trans}$, Eq.~(\ref{eq:ftrans}) (solid lines), and ${F}_{\rm trans}^\star$, Eq.~(\ref{eq:ftrans_eff}), with $\delta^{\rm f} = \gamma$ (dash-dotted lines).
As in the beta-equilibrium case discussed in Sect.~\ref{sec:result_betaequi}, both prescriptions, $F_{\rm trans}$ and ${F}_{\rm trans}^\star$, have the effect of reducing the cluster size. 
In particular, this effect is greater at lower $n_B$ and $Y_p^{\rm tot}$, and higher $T$.
However, in the scenario where the proton fraction is fixed, the impact of the translational free energy is only slightly reduced when the in-medium corrections on the cluster volume and effective mass are taken into account, as can be seen by comparing the solid and the dash-dotted lines in Fig.~\ref{fig:fixyp_Zvsnb_bsk24}. 

More strikingly, the role of the translational free energy is overall much less important than in the beta-equilibrium case. 
When the translational term is included, the clusters still survive up to around 0.1~fm$^{-3}$, unlike in beta equilibrium (see Table~\ref{table1:cc_transition} and Fig.~\ref{fig:betaequi_modifiedftrans_AZ}).
This can be understood by analysing the absolute ratio between the (ideal-gas) translational and the finite-size free energies, $|F_{\rm trans}/F_{\rm Coul + surf + curv}|$, shown in Fig.~\ref{fig:fixyp_fenergy_ratio_bsk24}.
Indeed, as already mentioned in Sect.~\ref{sec:result_betaequi}, the cluster size results from the competition between these two terms.
For the fixed proton fractions we consider, which are nevertheless higher than those obtained in beta equilibrium, the finite-size term dominates at all densities, and particularly at higher $n_B$, higher $Y_p^{\rm tot}$, and lower $T$, which explains the reduced impact of $F_{\rm trans}$ in these regimes observed in Fig.~\ref{fig:fixyp_Zvsnb_bsk24}. 
It is therefore natural to expect the effect of the translational free energy on the equation of state to remain negligible in the fixed-proton-fraction scenario as well, as can indeed be seen in Fig.~\ref{fig:fixyp_EoS_bsk24}.

\section{Conclusions}
\label{sec:conclusion}
In this work, we studied the influence of the translational degree of freedom in the liquid phase on the properties of finite-temperature ultra-dense stellar matter. 
To this aim, we derived different expressions for the effective mass of the ion moving in a uniform nucleon background using a hydrodynamical approach with different boundary conditions.
This renormalisation of the ion mass to account for the in-medium effects was then included in the translational-free-energy contribution of the ion energetics.
To calculate the latter, we employed a CLD model approach, in which the nuclear-matter properties were described using finite-temperature mean-field theory with nuclear energy-density functionals satisfying basic constraints from nuclear physics and astrophysics. Here, we optimised the surface parameters to reproduce the nuclear experimental masses consistently with the bulk energy. 
We performed this study at both beta equilibrium, as found in the proto-NS inner crusts close to the crystallisation transition, and at fixed proton fraction, which is relevant to the hot and dense  matter found in supernovae.

Using a standard variational procedure, we show that the size of the clusters is determined by the competition between the  motion of the centre-of-mass of the ion and the interface properties, namely, the Coulomb, surface, and curvature energies. 
Within the temperature range considered in this work (few $10^{10}$~K), the interface terms favour large clusters constituted of more than $\approx 100$ nucleons, while the entropy gain associated to the translational motion is maximal for a gas-like composition of very small ions.
Consequently, including the translational free energy in the variational theory can significantly reduce the number of nucleons in the clusters, especially at high temperatures. 

However, the importance of this effect varies with the scenario. 
At beta equilibrium, considering the ions as an ideal gas, the inclusion of the translational free energy leads to a drastic reduction of the cluster proton and mass numbers already at temperatures as low as the crystallisation temperature. 
This implies an early dissolution of clusters in the dense medium, hence a very low crust--core transition density. 
We believe that this is an important result, because the translational degrees of freedom are typically neglected in the literature when modelling the finite-temperature crust for NS cooling simulations \citep{Chamel2007, Fortin2010, Avancini2009, Avancini2017, Nandi2021}.
A quantitative estimation of the effect will require a fully quantal microscopic description of the dynamical properties of nuclei in a dense fermionic environment.
Such calculations are limited to the solid phase and moderate densities \citep{Baiko2001}, but could be in principle performed in the liquid phase similarly to in the work of \citet{Wlaz2016} and \citet{Sekizawa2022}, with adapted boundary conditions. 

Including the in-medium modification of the translational-free-energy expression resulting from considering that {only} neutrons in the continuum states participate in the flow, we obtain that, although the composition is strongly modified by the translational motion, the equation of state is just slightly affected.  
This unexpected influence of the translational entropy on the composition of matter is due to the fact that, in beta equilibrium, the proton fraction at high densities is very small, and the contribution from the Coulomb, surface, and curvature energies does not dominate the translational contribution.  
Conversely, if the total proton fraction is of the order of 0.2 or more, as typically found in supernova matter, the finite-size energies dominate, and the crust composition is less affected. 
Indeed, even at temperatures as high as  $k_{\rm B} T = 4$~MeV, heavy clusters are still favoured close to the transition to homogeneous matter. 

In supernova matter, neglecting the translational contribution in the variational treatment, as is done for instance in the popular equation of state by \citet{lattimer1991}, therefore appears to be justified. 
In this regime, the most questionable approximation is rather the OCP approximation, which we also used throughout this work. 
A comparison with a full MCP approach is in progress and will be presented in a separate paper.

\begin{acknowledgements}
This work has been partially supported by the IN2P3 Master Project NewMAC and the CNRS International Research Project (IRP) ``Origine des \'el\'ements lourds dans l’univers: Astres Compacts et Nucl\'eosynth\`ese (ACNu)''. 
We thank M. Antonelli for fruitful discussions.
\end{acknowledgements}

\end{document}